%% file: entel22.tex
\documentclass[12pt]{article}
\usepackage{amsmath}
\usepackage{amssymb}

\begin{document}

\title{\bf Entanglement and Teleportation\\
of Gaussian States of the Radiation Field}
\author{Paulina Marian$^1$,Tudor A. Marian$^2$, and Horia Scutaru$^3$\\ 
$^1$Department of Chemistry, University of 
Bucharest,\\
Blvd Regina Elisabeta 4-12, R-030018 Bucharest, Romania \\
$^2$Department of Physics, University of Bucharest, \\ P.O.Box MG-11,
R-077125 Bucharest-M\u{a}gurele, Romania\\
$^3$ Center for Advanced Studies in Physics of the Romanian Academy,
\\ 
Calea 13 Septembrie 13, R-050711 Bucharest, Romania}
\date{\em Received: December 19, 2002}
\maketitle
\begin{abstract}
We propose a reliable entanglement measure for a two-mode squeezed 
thermal state of the quantum electromagnetic field in terms of its 
Bures distance to the set of all separable states of the same kind.
The requisite Uhlmann fidelity of a pair of two-mode squeezed thermal 
states is evaluated as the maximal transition probability between two 
four-mode purifications. By applying the Peres-Simon criterion of 
separability we find the closest separable state. This enables us 
to derive an insightful expression of the amount of entanglement. 
Then we apply this measure of entanglement  to the study of the
Braunstein-Kimble protocol of teleportation. We use as input state 
in teleportation a mixed one-mode Gaussian state. The entangled state 
shared by the sender (Alice) and the receiver (Bob) is taken to be 
a two-mode squeezed thermal state. We find that the properties 
of the teleported state depend on both the input state and the 
 entanglement of the two-mode resource state. As a measure 
of the quality of the teleportation process, we employ the Uhlmann 
fidelity between the input and output mixed one-mode Gaussian
 states.
\end{abstract}


\section{Introduction} 

Most of the basic achievements in the rapidly developing field 
of quantum information theory have been obtained for finite-dimensional 
systems. However, ingenious protocols \cite{BK} and successful 
experiments \cite{Fur} reported in quantum teleportation of single-mode
states of the electromagnetic field justify our present interest 
in studying entanglement and teleportation of Gaussian field states. 

In the present work we review some recent progresses concerning the
fidelity of one-mode and two-mode Gaussian states and report our own 
results. We employ them to get an explicit expression 
of the {\it amount of entanglement} of a two-mode squeezed thermal 
state (STS). As an important application, teleportation of one-mode 
Gaussian states via a two-mode-STS channel is briefly discussed. 
We finally point out the properties of the fidelity of teleportation.
The paper is organized as follows. Section 2 is devoted 
to the description of the one-mode and two-mode Gaussian states 
of the quantum radiation field. Useful formulae for the fidelity 
of such states are presented in Sec. 3. In Sec. 4, we define
the degree of entanglement of a two-mode STS in terms of the Bures
distance between the state and the set of all separable STS's:
the resulting formula is at the same time simple and insightful. 
In Sec. 5, we describe by means of characteristic functions (CF's)
the teleportation of mixed one-mode states using as resource state
an entangled two-mode STS. For Gaussian states, the input-output
fidelity is analyzed as an appropriate measure 
of the efficiency of teleportation .

\section{Gaussian states}

\subsection{One-mode states}

Let
\begin{eqnarray}
a=\frac{1}{\sqrt{2}}(q+ip),\;\;\; a^{\dag}=\frac{1}{\sqrt{2}}(q-ip)
\end{eqnarray}
be the amplitude operators of the mode. Any single-mode Gaussian state 
is a displaced squeezed thermal state (DSTS):
\begin{eqnarray}
\rho=D(\alpha) S(r,\varphi) \rho_T S^{\dag}(r,\varphi)D^{\dag}(\alpha).
\end{eqnarray}
Here 
\begin{eqnarray}
D(\alpha):=\exp{(\alpha a^{\dag}-\alpha^* a)}
\end{eqnarray}
is a Weyl displacement operator with the coherent-state amplitude $\alpha 
\in \mathbb{C}$,
\begin{eqnarray}
S(r,\varphi):=\exp{\{\frac{1}{2} r[{\rm e}^{i\varphi} (a^{\dag})^2
-{\rm e}^{-i\varphi} a^2]\}}
\end{eqnarray}
is a Stoler squeeze operator with the squeeze factor  $r\geq 0$ 
and squeeze angle  $\varphi \in (-\pi,\pi]$, and
\begin{eqnarray}
\rho_{T}:=\frac{1}{\bar{n}+1}\sum_{n=0}^{\infty}
\left(\frac{\bar{n}}{\bar{n}+1}\right)^n|n \rangle \langle n|
\end{eqnarray}
is a Bose-Einstein density operator with the mean occupancy 
\begin{eqnarray}
\bar{n}=\left[\exp{\left(\frac{\hbar\omega}{k_BT}\right)}-1\right]^{-1}.
\end{eqnarray}
The Weyl expansion of the density operator,
\begin{eqnarray}
\rho=\frac{1}{\pi}\int {\rm d}^2 \lambda \;\;\chi (\lambda)
D(-\lambda),
\end{eqnarray}
with ${\rm d}^2 \lambda:={\rm d}\Re e(\lambda) {\rm d}\Im m (\lambda)$
points out the one-to-one correspondence between the field state $\rho$
and its CF
\begin{eqnarray}
\chi(\lambda):={\rm Tr}[\rho D(\lambda)].
\end{eqnarray}
By definition, a Gaussian state has a CF of the form
\begin{eqnarray}
\chi (\lambda)=\exp[-(A+\frac{1}{2})|\lambda|^2-
\frac{1}{2}B^{*}\lambda^2-\frac{1}{2}B(\lambda^{*})^2
+C^{*}\lambda-C\lambda^{*})],\; (A>0). 
\label{1mGCF} 
\end{eqnarray}
The coefficients $A,B,C$ in the exponent are determined 
by the DSTS parameters as
\begin{eqnarray}
A=\left(\bar{n}+\frac{1}{2}\right)\cosh (2r)
-\frac{1}{2},\;\;
B=-\left(\bar{n}+\frac{1}{2}\right) e^{i\varphi} \sinh(2r),\;\;
C=\alpha.
\end{eqnarray}
The covariance matrix,
\begin{eqnarray}
{\cal V}:=\left(\begin{array}{cc}\sigma(q,q)\;&\sigma(q,p)\\
\sigma(p,q)\;&\sigma(p,p) \end{array}\right), 
\label{2covmat}
\end{eqnarray}
allows one to write more compactly the generalized Heisenberg 
uncertainty relation,
\begin{eqnarray}  
{\rm det}\; {\cal V}\geq\frac{1}{4},
\end{eqnarray}
as well as the CF \ (\ref{1mGCF}):
\begin{eqnarray}
\chi (\lambda)=\exp{\{-\frac{1}{2} X^T {\cal V} X-i\;\Xi^TX\}}.
\label{1mGs}
\end{eqnarray}
We have denoted:
\begin{eqnarray}
\lambda:=-\frac{i}{\sqrt{2}}(x+iy),\;\;\;  X^T:=(x,y),
\end{eqnarray}
\begin{eqnarray}
\alpha:=\frac{1}{\sqrt{2}}(\xi+i\eta),\;\;\; \Xi^T:=(\xi,\eta).
\end{eqnarray}

\subsection{Nonclassicality} 
Classical states possess a well-behaved $P$ representation,
\begin{eqnarray}
\rho=\int{\rm d}^2 \alpha P(\alpha)|\alpha\rangle\langle\alpha|.
\end{eqnarray}
Otherwise, a state is {\em nonclassical}. For a Gaussian state, 
the integral
\begin{eqnarray} 
P(\alpha)=\frac{1}{\pi^2}\int {\rm d}^2 \lambda \exp(\alpha\lambda^*-
\alpha^*\lambda)\exp{(\frac{1}{2}|\lambda|^2)}\chi (\lambda)
\end{eqnarray}
exists if and only if $r\leq r_c$, where $r_c$ is 
the {\em nonclassicality threshold}, 
\begin{eqnarray} 
r_c:=\frac{1}{2}\ln (2\bar{n}+1).
\end{eqnarray}
Therefore, a Gaussian state is nonclassical if and only if $r>r_c.$
\subsection{Two-mode states}
We mention the Weyl expansion of a two-mode state:
\begin{eqnarray}
\rho=\frac{1}{\pi^2}\int {\rm d}^2 \lambda_1 {\rm d}^2 \lambda_2\;\;
\chi(\lambda_1,\lambda_2)D_1(-\lambda_1)D_2(-\lambda_2).
\end{eqnarray}
The CF of a Gaussian state has the explicit form
\begin{eqnarray}
\chi(\lambda_1,\lambda_2)=\chi_1(\lambda_1)\chi_2(\lambda_2)
\exp{[- F\lambda_1^*\lambda_2-
 F^*\lambda_1\lambda_2^*+G^*\lambda_1\lambda_2+G\lambda_1^*\lambda_2^*]}.
\label{2mGFC}
\end{eqnarray}
The formula similar to Eq.\ (\ref{1mGs})is
\begin{eqnarray}
\chi(\lambda_1,\lambda_2)=\exp{\{-\frac{1}{2}X^T {\cal V} X -i\;\Xi^T X\}}
\label{2mGs}
\end{eqnarray}
with
\begin{eqnarray}
X^T=(x_1,y_1,x_2,y_2),\;\;\;\Xi^T=(\xi_1,\eta_1,\xi_2,\eta_2).
\end{eqnarray}
In Eq.\ (\ref{2mGs}), ${\cal V}$ is the real, symmetric, and positive
$4\times 4$ covariance matrix
\begin{eqnarray}
{\cal V}=\left(\begin{array}{cc}{\cal V}_1\;\;&{\cal C}\\
{\cal C}^T\;\;&{\cal V}_2\end{array}\right),
\label{4covmat}
\end{eqnarray}
where ${\cal V}_j, \;\; (j=1,2)$ are $2\times 2$
single-mode reduced covariance matrices of the form \ (\ref{2covmat}),
and {${\cal C}$} is the cross-covariance matrix,
\begin{eqnarray}
{\cal C}=\left(\begin{array}{cc} \sigma (q_1, q_2)\;\;&\sigma (q_1, p_2)\\
 \sigma (p_1, q_2)\;\;&\sigma (p_1, p_2)\end{array}\right).
\end{eqnarray}
There are four independent invariants under local symplectic 
transformations Sp$(2,\mathbb{R})\; \otimes $ Sp$(2,\mathbb{R})$:
$\rm det{\cal V}_1, \;\; \rm det{\cal V}_2, \;\;
\rm det{\cal C}, \;\; \rm det{\cal V}.$
An inequality that incorporates the Heisenberg uncertainty relations
can be expressed in terms of them as
\begin{eqnarray}
{\rm det}{\cal V}-\frac{1}{4}\left[{\rm det}{\cal V}_1
+{\rm det}{\cal V}_2+2 {\rm det}{\cal C}\right]+\frac{1}{16}\geq 0.
\end{eqnarray}
An important class of mixed Gaussian states consists of two-mode STS's.
Such a state is the unitary transform of a two-mode thermal state, 
\begin{eqnarray}
\rho=S_{12}(r, \varphi)(\rho_{T_1}\otimes\rho_{T_2})
S_{12}^{\dag}(r, \varphi),
\label{2STS}
\end{eqnarray}
by a two-mode squeeze operator,
\begin{eqnarray}
S_{12}(r, \varphi):=\exp{[r({\rm e}^{i\varphi} 
a_1^{\dag}a_2^{\dag}-{\rm e}^{-i\varphi} a_1a_2)]}.
\end{eqnarray}
A two-mode STS \ (\ref{2STS}) can be experimentally prepared 
by parametric amplification of light. Its local invariants read:
{\begin{eqnarray}
\bar{N}_{1,2}+\frac{1}{2}:=\sqrt{\rm det {\cal V}_{1,2}}=
\left(\bar{n}_{1,2}+\frac{1}{2}\right)(\cosh r)^2+
\left(\bar{n}_{2,1}+\frac{1}{2}\right)(\sinh r)^2,
\end{eqnarray}
\begin{eqnarray}
\sqrt{-\rm det {\cal C}}=(\bar{n}_1+\bar{n}_2+1)\sinh r\cosh r,
\end{eqnarray}
\begin{eqnarray}
\sqrt{\rm det {\cal V}}=\left(\bar{n}_1+\frac{1}{2}\right)
\left(\bar{n}_2+\frac{1}{2}\right).
\end{eqnarray}

\section{Fidelity}
\subsection{General properties}
\setcounter{equation}{0}
Pure states of a quantum mechanical system are described 
by unit rays in the Hilbert space,
\begin{eqnarray}
f=\{{\rm e}^{i\theta}|\Psi\rangle\} 
\end{eqnarray}
The squared distance between
two unit rays is
\begin{eqnarray}
[d(f_1,f_2)]^2&=&
\min|||\Psi_1\rangle- {\rm e}^{i\theta}|\Psi_2\rangle||^2
\nonumber\\&& =2(1-|\langle \Psi_1 |\Psi_2\rangle|)
\end{eqnarray}
Consider now a mixed state $\rho$ of a quantum system 
whose Hilbert space is ${\cal H}_A$. A {\em purification} 
of $\rho$ is a pure state $|\Phi\rangle \langle\Phi|$ 
on a tensor product of Hilbert spaces ${\cal H}_A \otimes {\cal H}_B$ 
such that its reduction to ${\cal H}_A$ is the given mixed state:
\begin{eqnarray}
\rho={\rm Tr}_B (|\Phi\rangle \langle\Phi|).
\end{eqnarray}
To a pair of mixed states on ${\cal H}_A$, $\rho_{1,2}$, 
one can associate a pair of purifications,
$|\Phi_{1,2}\rangle \langle\Phi_{1,2}|$, on 
${\cal H}_A \otimes {\cal H}_B$, which is not unique. 
The squared Bures distance between $\rho_{1}$ and $\rho_{2}$, 
originally introduced on mathematical grounds \cite{Bures}, is
\begin{eqnarray}
d_{B}^2(\rho_1,\rho_2):=\min|||\Phi_1\rangle-|\Phi_2\rangle||^2
=2(1-\max|\langle \Phi_1 |\Phi_2\rangle|).
\label{bures}
\end{eqnarray}
Since this definition can obviously be extended to pure states,
the set of all quantum states (pure and mixed) may be equipped 
with the Bures distance to become a metric space. Notice that
the "transition probability" between the mixed states $\rho_1$ 
and $\rho_2$, defined later by Uhlmann \cite{Uhl}, 
\begin{eqnarray}
{\cal F}(\rho_1,\rho_2):=\text{max}|\langle \Phi_1|\Phi_2 \rangle|^2,
\label{fidpur}
\end{eqnarray}
is closely related to the Bures metric: 
\begin{eqnarray}
d_{B}(\rho_1,\rho_2)=\sqrt{2-2\sqrt{{\cal F}(\rho_1,\rho_2)}}.
\end{eqnarray}
Uhlmann \cite{Uhl} succeeded in deriving an intrinsic expression 
of the quantity \ (\ref{fidpur}), now called {\em fidelity} \cite{Jozsa}: 
\begin{eqnarray}
{\cal F}(\rho_1,\rho_2)=\left\{Tr[(\sqrt{\rho_1}\rho_2
\sqrt{\rho_1})^{1/2}]\right\}^2.
\end{eqnarray}
We list some properties of the fidelity 
\cite{Uhl,Jozsa,BC,Fuchs,Barnum,Nielsen}:
\begin{enumerate}
\item { $0\leq {\cal F}(\rho_1,\rho_2)\leq 1$}, 
and {${\cal F}(\rho_1,\rho_2)=1$} if and only if 
{$\rho_1=\rho_2$};
\item {${\cal F}(\rho_1,\rho_2)
={\cal F}(\rho_2,\rho_1)$}, \;\;\; (symmetry);
\item {${\cal F}(\rho_1,\rho_2)
\geq {\rm Tr}(\rho_1\rho_2)$;}  if {$\rho_1$} 
or/and {$\rho_2$} are pure, 
then {${\cal F}(\rho_1,\rho_2)=
{\rm Tr}(\rho_1\rho_2)$};
\item {${\cal F}(\rho_1\otimes \sigma_1,\rho_2 
\otimes \sigma_2)=
{\cal F}(\rho_1,\rho_2){\cal F}(\sigma_1,\sigma_2)$}, 
\;\;\;(multiplicativity);
\item{ ${\cal F}(U\rho_1 U^{\dag},U\rho_2U^{\dag})
={\cal F}(\rho_1,\rho_2)$}, \;\;\;(invariance under unitary 
transformations);
\item { $\sqrt{{\cal F}(\rho_1,\rho_2)}
=\min_{ \{E_b\} } \sum_{b}
\sqrt{{\rm Tr} (\rho_1 E_b)}
\sqrt{{\rm Tr} (\rho_2 E_b)},$ }\;\;\; where $\{E_b\}$ is any set 
of nonnegative operators which is complete, 
{\em i. e.} 
$\sum_b{E_b}=I;$ such a set $\{E_b\}$ is called a positive 
operator-valued measure (POVM) and 
$p_{1,2}(b)={\rm Tr}(\rho_{1,2} E_b)$ are probability distributions 
generated by it.
\end{enumerate}

\subsection{One-mode Gaussian states}

There are few quantum systems for which an explicit expression 
for the fidelity of two mixed states is available so far. 
An important formula has been recently established 
for one-mode Gaussian states of the radiation field: 
first Twamley \cite{Twam} obtained the fidelity of two STS's, 
and later Scutaru \cite{Horia} derived the expression 
for the fidelity of any pair of DSTS's. The latter formula reads:
\begin{eqnarray}
&&F(\rho,\rho^{\prime})
=\left(\sqrt{\Delta+\Lambda}-\sqrt{\Lambda}\right)^{-1}
\nonumber\\ &&\times \exp{\left[-\frac{1}{\Delta}\left[
(A+A^{\prime}+1)|C-C^{\prime}|^2
-\Re e[(B+B^{\prime})
(C^*-C^{\prime *})^2\right]\right]},
\label{1fid}
\end{eqnarray}
with
\begin{eqnarray}
\Delta:={\rm det}({\cal V}+{\cal V}^{\prime}),\;\;
\Lambda:=4\left[{\rm det}({\cal V})-\frac{1}{4}\right]
\left[{\rm det}({\cal V}^{\prime})-\frac{1}{4}\right].
\label{DeLa}
\end{eqnarray}

\subsection{Degree of nonclassicality}

As an application of the previous formula, Eqs.\ (\ref{1fid}) and
\ (\ref{DeLa}), we evaluated the {\em degree of nonclassicality}
 of a single-mode Gaussian state $\rho$ \cite{PTH02}. We defined
this quantity in terms of the Bures distance between the state $\rho$ 
and the set {${\cal C}_0$} of all classical one-mode Gaussian states:
\begin{eqnarray}
Q_0(\rho):=\frac{1}{2} \min_{{\rho}^{\prime} \in {\cal C}_0}
d^2_B(\rho, {\rho}^{\prime}).
\label{degQ}
\end{eqnarray}
We established the result 
\begin{eqnarray}
Q_0(\rho)=0,\;\;\;(r\leq r_c),
\end{eqnarray}
\begin{eqnarray}
Q_0(\rho)=1-[{\rm sech} (r-r_c)]^{1/2},  \;\;\; (r>r_c),
\label{q}
\end{eqnarray}
which fulfils the following three requirements:\\
Q1) $Q_0(\rho)$ vanishes if and only if the state $\rho$ 
is classical;\\
Q2) Classical transformations (defined as mapping coherent states
into coherent states) preserve  $Q_0(\rho)$;\\
Q3) Nonclassicality does not increase under any POVM mapping.

\subsection{The Schmidt decomposition}

A  pure bipartite state of a composite system is called {\em separable} 
or {\em disentangled} when its state vector is a direct product 
of two state vectors of the parts:
\begin{eqnarray}
|\Psi\rangle=|\Psi_A\rangle\otimes|\Psi_B\rangle, 
\label{prod}
\end{eqnarray}
where $|\Psi_A\rangle \in {\cal H}_A$ and $|\Psi_B\rangle 
\in {\cal H}_B$. In the opposite case of an {\em inseparable} 
or
{\em entangled} pure bipartite state, we point out 
a very useful biorthogonal expansion of the state vector,
which is due to Erhard Schmidt \cite {Schmidt}:
\begin{eqnarray}
|\Psi \rangle=\sum_{n=1}^{N}\sqrt{\lambda_n}
|\Psi_{A}^{(n)} \rangle \otimes |\Psi_{B}^{(n)} \rangle, 
\label{Sch} 
\end{eqnarray}
with {$\sum_{n=1}^{N}\lambda_n=1$}. The squared Schmidt 
coefficients $\lambda_n$ are the common positive eigenvalues 
of the reductions $\rho_{A,B}:=Tr_{B,A}(|\Psi \rangle \langle \Psi|)$.
The corresponding eigenvectors, $|\Psi_{A}^{(n)}\rangle$ 
in ${\cal H}_A$ and $|\Psi_{B}^{(n)}\rangle$ in ${\cal H}_B$, belong
to the Schmidt bases in these spaces. Accordingly, the {\em Schmidt rank} 
$N$ cannot exceed the dimensions of the factor Hilbert spaces: \;\;\;
$2\leq N\leq \min(d_1,d_2)$. The value $N=1$ is excluded because 
it corresponds to the product state, Eq.\ (\ref{prod}).
By definition, $|\Psi \rangle $ is a purification of both $\rho_{A}$ 
and $\rho_{B}$. 

We are interested in the converse problem: 
Starting from a mixed state ${\rho_A} $ in ${\cal H}_A$, 
construct purifications in an extended Hilbert space 
${\cal H}={\cal H}_A\otimes{\cal H}_A$ by using 
the eigenvalue problem of ${\rho_A}$ and the Schmidt decomposition.

\subsection{Two-mode STS's}

We have evaluated the fidelity of a pair of two-mode STS's by applying
Eq.\ (\ref{fidpur}) \cite{PT}. The eigenvalues of the density operator
\ (\ref{2STS}) are  
\begin{eqnarray}
\lambda_{kl}=\frac{(\bar{n}_1)^k (\bar{n}_2)^l}{(\bar{n}_1+1)^{k+1}
(\bar{n}_2+1)^{l+1}}.
\end{eqnarray}
To a pair of two-mode STS's $\rho$ (parameters: $\bar{n}_1,\bar{n}_2, 
r, \varphi$) and $\rho^{\prime}$ (parameters: $\bar{n}_1^{\prime},
\bar{n}_2^{\prime},r^{\prime}, \varphi^{\prime}$) we associate 
the following pair of four-mode purifications written as Schmidt series,
 Eq.\ (\ref{Sch}):
\begin{eqnarray}
|\Psi \rangle=\sum_{kl} \sqrt{\lambda_{kl}} S_{12}(r, \varphi)|kl \rangle
\otimes |kl \rangle
\label{psi}
\end{eqnarray}
and
\begin{eqnarray}
|\Psi^{\prime} \rangle=\sum_{mn} \sqrt{\lambda_{mn}^{\prime}} 
S_{12}(r^{\prime}, \varphi^{\prime})|mn \rangle \otimes {U}|mn \rangle.
\label{psipr}
\end{eqnarray}
In Eq.\ (\ref{psipr}), $U$ is a unitary operator,
\begin{eqnarray}
U:= {\rm e}^{-i\vartheta}R_{12}(\vartheta)S_{12}(\varrho, \phi),
\label{unit}
\end{eqnarray}
whose second factor is a two-mode rotation operator,
\begin{eqnarray}
R_{12}(\vartheta):=\exp{[-i\vartheta(a_1^{\dag}a_1+a_2^{\dag}a_2)]}.
\end{eqnarray}
$U$ depends on three free variables: the rotation angle $\vartheta$,
the squeeze factor $\varrho$, and the squeeze angle $\phi$. We obtained 
a compact form of the quantum mechanical transition probability between
the purifications \ (\ref{psi}) and \ (\ref{psipr}). Its maximum value
with respect to the parameters $\vartheta, \varrho, \phi$  yields,
according to Eq.\ (\ref{fidpur}), the fidelity of two arbitrary
two-mode STS's:
\begin{eqnarray}
{\cal F}(\rho, \rho^{\prime})=\{[\sqrt{\rm det ({\cal V}+
{\cal V}^{\prime})}+(\sqrt{X_{1}}+\sqrt{X_{2}})^2]^{1/2}
-\sqrt{X_{1}}-\sqrt{X_{2}}\}^{-2},
\label{2fid}
\end{eqnarray}
where 
\begin{eqnarray}
X_{1,2}:=\bar{n}_{1,2}\bar{n}_{1,2}^{\prime}
(\bar{n}_{2,1}+1)(\bar{n}_{2,1}^{\prime}+1).
\end{eqnarray} 

\section{Entanglement}
\setcounter{equation}{0}
\subsection{Inseparable quantum states}

In order to include the mixed-state case, Werner \cite{Werner} 
gave the following general definition of the separability. 
A separable (or disentangled) bipartite state $\rho_{AB}$ is a convex 
combination of product states $\rho_A^{(i)}\otimes\rho_B^{(i)}$:
\begin{eqnarray}
\rho_{AB}=\sum_{i}p_i \rho_A^{(i)}\otimes\rho_B^{(i)},
\;\;\;p_i\geq 0,\;\;\;\sum_{i}p_i=1.
\label{sep}
\end{eqnarray}
If the state is not such a mixture, it is termed inseparable 
(or entangled).
For instance, a classical two-mode state 
of the radiation field is separable, but the converse is not true.

There are two main open problems concerning the entanglement 
of mixed states.
They refer to:
\begin{itemize} 
\item separability criteria: no universal criterion was still formulated;
\item measures of entanglement: no universal entanglement measure 
could be applied. 
\end{itemize}
 
\subsection{The Peres condition of separability} 
 
Peres \cite{Peres} found a general necessary condition 
for the separability of a bipartite state: this is the preservation 
of the nonnegativity of the density matrix under partial transposition.
This condition is not a universally sufficient one. However, Horodecki
\cite{Hor} and Simon \cite{Simon} proved that, in two important cases, 
Peres' statement is also a criterion for separability. These are,
respectively:
\begin{description}

\item a) two-spin-${\frac{1}{2}}$ states and also
spin-${\frac{1}{2}}$-spin-1 states;
\item b) two-mode Gaussian states of the radiation field.
\end{description}

In the latter case, Simon \cite{Simon} gave a Sp$(2,\mathbb{R})\; \otimes $
Sp$(2,\mathbb{R})$ invariant form of the separability criterion:
\begin{eqnarray}
{\rm det}{\cal V}-\frac{1}{4}
\left[{\rm det}{\cal V}_1+{\rm det}{\cal V}_2+2|{\rm det}{\cal C}|
\right]+\frac{1}{16}\geq 0.
\label{Sim}
\end{eqnarray}
Using Eq.\ (\ref{Sim}), one can easily check whether a two-mode Gaussian 
state is separable or not \cite{PTH01}. In particular, for two-mode STS's, 
the Peres-Simon criterion \ (\ref{Sim}) reads:
\begin{eqnarray}
(\cosh r)^2\leq (\cosh r_s)^2:=\frac{(\bar{n}_1+1)(\bar{n}_2+1)}
{\bar{n}_1+\bar{n}_2+1}.
\label{SimSTS}
\end{eqnarray}

\subsection{Measures of entanglement}

In their search for a good measure of entanglement, $E(\rho)$,
of a bipartite state $\rho$, Vedral {\em et al} stated 
in Ref. \cite{Ved} the following general demands:\\
E1) $E(\rho)$ vanishes if and only if the state $\rho$ is separable;\\
E2) local unitary transformations preserve $E(\rho)$;\\
E3) $E(\rho)$ does not increase under local general measurements.\\
They proved that a convenient entanglement measure could be 
a distance between the state $\rho$ and the set {${\cal D}$} 
of all the separable states of the given system:
\begin{eqnarray}
E(\rho):=\min_{\sigma \in {\cal D}}d(\rho, \sigma).
\label{ent}
\end{eqnarray}
Two candidates for the distance $d(\rho, \sigma)$ in Eq.\ (\ref{ent}) 
are found to be suitable:
\begin{itemize}
\item The quantum relative entropy,
\begin{eqnarray}
S(\sigma||\rho):=Tr [\sigma(\ln\sigma -\ln \rho)],
\label{relent}
\end{eqnarray}
which is not a true metric;
\item The Bures metric, Eq.\ (\ref{bures}).
\end{itemize} 
Vedral {\em et al} \cite{Ved} succeeded in obtaining an explicit
expression of the amount of entanglement $E(\rho)$ 
of any {\em pure} bipartite state $\rho$, by choosing as "distance" 
in  Eq.\ (\ref{ent}) the relative entropy, Eq.\ (\ref{relent}). 
For $d(\rho, \sigma)=S(\sigma||\rho)$ they found that the measure
of its entanglement is the common von Neumann entropy 
of the reduced states $\rho_{A}$ and $\rho_{B}$:
\begin{eqnarray}
E(\rho)=S(\rho_{A}):=-Tr_A[\rho_{A}\ln(\rho_{A})].
\label{JvN}
\end{eqnarray}

\subsection{Entanglement of a two-mode STS}

In order to estimate the amount of entanglement of a two-mode STS, 
we apply Eq.\ (\ref{ent}) using the Bures distance. However, 
we make an upper bound approximation by replacing the set {${\cal D}$} 
of all the separable two-mode states by its subset ${\cal D}_0$
consisting of all the separable STS's. Hence for a given 
inseparable two-mode STS, $\rho$, we evaluate the following 
degree of entanglement:  
\begin{eqnarray}
E_0(\rho):=\min_{{\rho}^{\prime} \in {\cal D}_0}\frac{1}{2} 
d^2_B(\rho, {\rho}^{\prime})=1-\max_{{\rho}^{\prime} \in {\cal D}_0}
\sqrt{{\cal F}(\rho, {\rho}^{\prime})}.
\end{eqnarray} 
We make use of the fidelity \ (\ref{2fid}) of the given inseparable
STS $\rho$ $(\text{parameters}: \bar{n}_1,\bar{n}_2, r>r_s, \varphi)$ 
with respect to an arbitrary separable STS $\rho^{\prime}\in {\cal D}_0$}

$(\text{parameters}: \bar{n}_1^{\prime},\bar{n}_2^{\prime},
r^{\prime},\varphi^{\prime},\;\; \text{with}\;\; 
r^{\prime}\leq {r_s}^{\prime})$. 
We determine the parameters of the closest separable STS $\tilde{\rho}$:
$\tilde{\varphi}=\varphi,\; \tilde{r}=\tilde{r}_s,\;
\tilde{\bar{n}}_{1},\; \tilde{\bar{n}}_{2}$,
and then get a significant formula \cite{PTH03},
\begin{eqnarray}
E_0(\rho)=0,\;\;\;(r\leq r_s),
\end{eqnarray}
\begin{eqnarray} 
E_0(\rho)=1-{\rm sech} (r-r_s),\;\;\;(r>r_s),
\label{entSTS}
\end{eqnarray}
which observes the demands E1)-E3) for an adequate measure 
of entanglement.

\section{Teleportation}
\setcounter{equation}{0}
\subsection{Spin-${\frac{1}{2}}$ states} 

The discovery of the possibility of quantum teleportation 
by Bennett {\em et al} \cite{Bennett} opened new research directions 
in the field of quantum processing of information. We quote from 
Ref.\cite{Fur}: "Quantum teleportation is the disembodied transport
of an unknown quantum state from one place to another". The key idea
of the {\em Gedankenexperiment} described in Ref.\cite{Bennett} 
is that two distant operators, Alice at a sending station and Bob 
at a receiving terminal, share an entangled quantum bipartite state 
and exploit its {\em nonlocal} character as a quantum resource.
The resource state, which is also called an Einstein-Podolsky-Rosen 
(EPR) state \cite{EPR}, is here the singlet state of a pair 
of spin-${\frac{1}{2}}$ particles. Particle 1 is given to Alice and 
particle 2 is given to Bob. Alice intends to transport an {\em unknown} 
state of a third spin-${\frac{1}{2}}$ particle to Bob. She performs 
a complete projective measurement on the joint system 
and then conveys its outcome to Bob via a classical 
communication channel. As a consequence of Alice's measurement, 
the total-spin state of the three-particle system collapses. Due to the 
entanglement, this involves a breakdown of the spin-${\frac{1}{2}}$ state 
of Bob's particle 2.  Nevertheless, Bob makes use of the information 
transmitted classically by Alice to transform his reduced state into 
an output that is an accurate replica of the original unknown input.

\subsection{One-mode states of the radiation field}

Along the lines sketched above, Braunstein and Kimble \cite{BK}
put forward a teleportation protocol for optical one-mode field states. 
They propose as resource state shared by Alice and Bob a two-mode 
squeezed vacuum state (SVS). Very soon, this protocol was implemented 
into a successful experiment that demonstrated the quantum teleportation 
of optical coherent states \cite{Fur}. It is useful to present briefly 
the Braunstein-Kimble protocol. It consists of three steps, as follows.\\ 
{$\bullet$ Step 1}. Alice mixes two waves with a beam splitter, 
namely an unknown one-mode input in the state $\rho_{in}$ and
 the two-mode beam in the  EPR state $\rho_{AB}$.\\
{$\bullet$ Step 2}. Alice measures simultaneously the observables
 $ q= q_{in}-q_A,\;\;  p=p_{in}+p_A$ in the resulting three-mode state.
Quantum Mechanics predicts their distribution function:
\begin{eqnarray}
{\cal P}(q,p)={\rm Tr}_{in,AB}\left[P(\rho_{in}
\otimes\rho_{AB})P^{\dag}\right]
\end{eqnarray}
with
\begin{eqnarray}
P=|\Phi_{in,A}(q,p)\rangle
\langle\Phi_{in,A}(q,p)|\otimes I_B.
\end{eqnarray}
The complete von Neumann measurement performed by Alice entails 
a collapse of the tripartite state to a state whose reduction 
at Bob's disposal is
\begin{eqnarray}
\rho_B^{\prime}=\frac{1}{{\cal P}(q,p)}{\rm Tr}_{in,A}
\left[P(\rho_{in}\otimes\rho_{AB})P^{\dag}\right].
\end{eqnarray}
{$\bullet$ Step 3}. Using classical communication lines, Alice conveys 
to Bob the outcome $\{q, p\}$ of her measurement. Then, Bob superposes 
a coherent field whose amplitude is precisely 
{$\mu=\frac{1}{\sqrt{2}}(q+ip)$ on the mode ${\rho}_B^{\prime}$ 
at his hand:
\begin{eqnarray}
\rho_B^{\prime}\longrightarrow \rho_B^{ \prime \prime}=D_2(\mu)
\rho_B^{\prime}D_2^{\dag}(\mu).
\end{eqnarray}
A more realistic ensemble description of the projective measurements 
carried out by Alice yields the teleported state
\begin{eqnarray}
\rho_{out}=\int\int {\rm d}q{\rm d}p{\cal P}(q,p)\rho_B^{\prime \prime}.
\end{eqnarray}

\subsection{CF of the teleported state}

The common eigenfunction of the pair of continuous quantum variables 
$\{q, p\}$ measured by Alice,
\begin{eqnarray}
&&|\Phi_{in,A}(q,p)\rangle=\frac{1}{\sqrt{2 \pi}}
\int\limits_{-\infty}^{\infty}{\rm d}\eta{\rm e}^{ip\eta}
|q+\eta\rangle_{in}\otimes|\eta\rangle_A,
\end{eqnarray}
has the coherent-state expansion
\begin{eqnarray}
&&|\Phi_{in,A}(q,p)\rangle=\frac{1}{\sqrt{2\pi}}
\exp\left[-\frac{|\mu|^2}{2}-\frac{ipq}{2}\right] 
\frac{1}{\pi^2}\int
\int{\rm d}^2 \alpha \;{\rm d}^2 \beta \;|\alpha\rangle_{in} 
\otimes| \beta\rangle_{A}\nonumber\\ && \times
\exp{\left[-\frac{|\alpha|^2}{2}-\frac{|\beta|^2}{2}
+\alpha^*\beta^*-\mu^*\beta^*+\mu \alpha^*\right]}.
\end{eqnarray}
Using the Weyl expansions of the density operators $\rho_{in}, \rho_{AB},
\rho^{\prime}_{B}, \rho_{out},$ we get the CF of the state at Bob's hand 
after the phase-space translation performed by him:
\begin{eqnarray}
&&\chi_{out}(\lambda)= \chi_{in}(\lambda)\chi_{AB}(\lambda^*,\lambda).
\label{telCF}
\end{eqnarray}
Equation \ (\ref{telCF}) is the main result of the present work. 
Therefore, if the EPR state ${\rho}_{AB}$ is a two-mode Gaussian state, 
then any single-mode Gaussian input is teleported as a single-mode 
Gaussian output.

In what follows we choose as EPR state an entangled two-mode 
{\em symmetric} STS. This is a state \ (\ref{2STS}) having equal 
mode frequencies and possessing equal mean numbers of thermal photons 
in both modes, $\bar{n}_{1}=\bar{n}_{2}<{\frac{1}{2}}(e^{2r}-1)$; 
in addition, we assume that the squeeze angle is equal to zero:
\begin{eqnarray}
\rho_{AB}=S_{12}(r,0)(\rho_T\otimes\rho_T)S_{12}^{\dag}(r,0).
\label{EPR2STS} 
\end{eqnarray}
Then Eq. \ (\ref{telCF}) shows that teleportation of a Gaussian state 
merely provides additional noise, exactly like superposition 
of a thermal field:
\begin{eqnarray}
A_{out}=A_{in}+ \exp{[-2(r-r_s)]},\;\;\; B_{out}=B_{in},\;\;\; 
C_{out}=C_{in}.
\end{eqnarray}
 
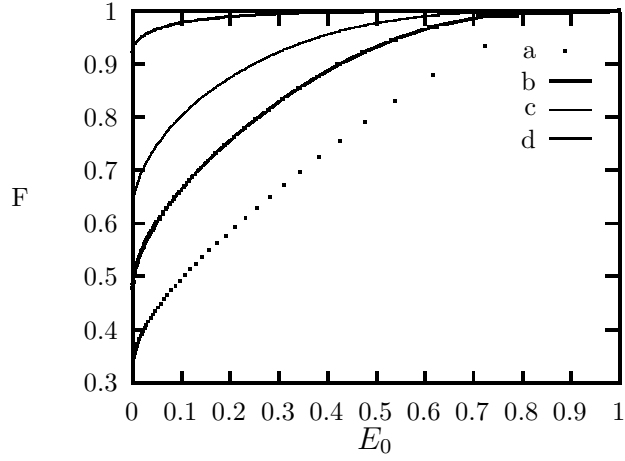
\begin{figure}
\begin{center}
\input{nec.tex}
\caption{ Fidelity  of teleportation for several mixed Gaussian states  
versus the entanglement of the EPR state \ (\ref{EPR2STS}). 
The squeeze factor is $r=1$. The curves a ($\bar{n}=0$), 
b ($\bar{n}=0.1$), c ($\bar{n}=0.5$) correspond to nonclassical input 
states while the d one is for the classical state with ($\bar{n}=5$). 
It is clear that the fidelity increases with the degree of mixing.}
\end{center}
\end{figure}
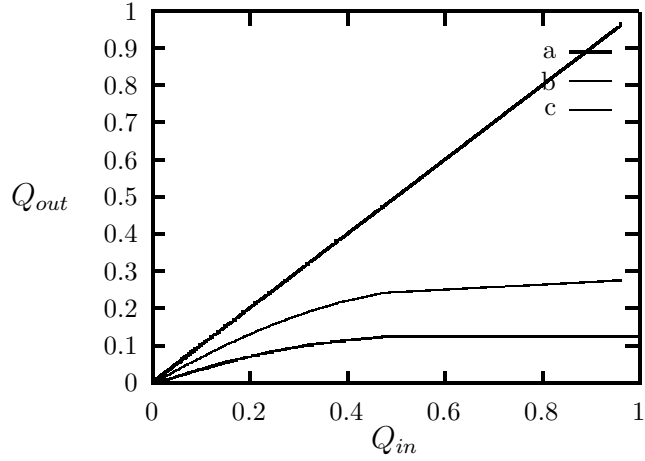
\begin{figure}
\begin{center}
\input{non.tex}
\caption{Degree of nonclassicality of the teleported
Gaussian state  versus the similar quantity of the
input state plotted for several values of the entanglement 
of the EPR state \ (\ref{EPR2STS}) : $E_0=1$ (curve a),$E_0=0.615$ 
(curve b), $E_0=0.425 $ (curve c). Both measures of nonclassicality  
and entanglement are defined using the Bures distance, 
Eqs.\ (\ref{q}) and \ (\ref{entSTS}), respectively.}
\end{center}
\end{figure}

\subsection{Fidelity of teleportation}

The quality of the teleportation protocol is quantified 
by the output-input fidelity, called fidelity of teleportation.
In particular, according to Eqs.\ (\ref{1fid}) and \ (\ref{DeLa}), 
the fidelity of teleportation of pure or mixed one-mode Gaussian states
via the EPR state \ (\ref{EPR2STS}) is
\begin{eqnarray}
{\cal F}(\rho_{out},\rho_{in})
=\left(\sqrt{\Delta+\Lambda}-\sqrt{\Lambda}\right)^{-1},
\end{eqnarray}
where
\begin{eqnarray}
\Delta=4\left(y^2+xyz+\frac{1}{4}z^2\right),\;\;
\Lambda=4\left(y^2-\frac{1}{4}\right)\left(y^2-\frac{1}{4}
+2xyz+z^2\right),
\end{eqnarray}
with the variables
\begin{eqnarray}
 x:=\cosh (2r_{in})\geq 1,\;\;
y:=\bar{n}_{in}+\frac{1}{2} \geq \frac{1}{2},\;\;
z:=\exp{[-2(r-r_s)]>0}.
\end{eqnarray}
Note that the parameters $x$ and $y$ are characteristics of the input 
one-mode DSTS, while $z$ depends only on the degree of entanglement, 
Eq.\ (\ref{entSTS}), of the EPR state \ (\ref{EPR2STS}).

We mention some properties of the fidelity of teleportation
\begin{eqnarray}
{\cal F}(\rho_{out},\rho_{in}):={\cal F}(x,y,z).
\end{eqnarray}
\begin{itemize}
\item
$\frac{\partial {\cal F}}{\partial x}<0 $: 
Fidelity decreases with the squeeze factor of the input state.
\item
$\frac{\partial {\cal F}}{\partial y}>0 $: 
Fidelity decreases with the degree of purity of the input state.
\end{itemize}
Accordingly, for $r_{in}>r_c$, fidelity decreases with 
the degree of nonclassicality.
\begin{itemize}
\item
$\frac{\partial {\cal F}}{\partial z}<0 $:
Fidelity increases with the entanglement of the STS resource.
\end{itemize}
Indeed,
\begin{eqnarray} 
E_0({\rho}_{AB})=\frac{(1-\sqrt{z})^2}{1+z}, 
\;\;\; (z<1 \Longleftrightarrow r>r_{s}).
\end{eqnarray}
In the special case of an input coherent state, we recover the formula
\cite{Ide}
\begin{eqnarray}
{\cal F}(1,\frac{1}{2},z)=\frac{1}{1+z},
\end{eqnarray}
so that
\begin{eqnarray}
{\cal F}(1,\frac{1}{2},z)>\frac{N}{N+1} \Longleftrightarrow 
r>r_{s}+\frac{1}{2}\ln{N},\;\;\;\; (N=1,2,3,...).
\end{eqnarray}
\section*{Acknowledgment}
This work was supported by the Romanian 
CNCSIS through a grant for the University of Bucharest.

\end{document}

%% file: nec.tex
\setlength{\unitlength}{0.240900pt}
\ifx\plotpoint\undefined\newsavebox{\plotpoint}\fi
\sbox{\plotpoint}{\rule[-0.500pt]{1.000pt}{1.000pt}}%
\begin{picture}(1049,720)(0,0)
\font\gnuplot=cmr10 at 10pt
\gnuplot
\sbox{\plotpoint}{\rule[-0.500pt]{1.000pt}{1.000pt}}%
\put(220.0,113.0){\rule[-0.500pt]{1.000pt}{140.686pt}}
\put(220.0,113.0){\rule[-0.500pt]{4.818pt}{1.000pt}}
\put(198,113){\makebox(0,0)[r]{0.3}}
\put(965.0,113.0){\rule[-0.500pt]{4.818pt}{1.000pt}}
\put(220.0,196.0){\rule[-0.500pt]{4.818pt}{1.000pt}}
\put(198,196){\makebox(0,0)[r]{0.4}}
\put(965.0,196.0){\rule[-0.500pt]{4.818pt}{1.000pt}}
\put(220.0,280.0){\rule[-0.500pt]{4.818pt}{1.000pt}}
\put(198,280){\makebox(0,0)[r]{0.5}}
\put(965.0,280.0){\rule[-0.500pt]{4.818pt}{1.000pt}}
\put(220.0,363.0){\rule[-0.500pt]{4.818pt}{1.000pt}}
\put(198,363){\makebox(0,0)[r]{0.6}}
\put(965.0,363.0){\rule[-0.500pt]{4.818pt}{1.000pt}}
\put(220.0,447.0){\rule[-0.500pt]{4.818pt}{1.000pt}}
\put(198,447){\makebox(0,0)[r]{0.7}}
\put(965.0,447.0){\rule[-0.500pt]{4.818pt}{1.000pt}}
\put(220.0,530.0){\rule[-0.500pt]{4.818pt}{1.000pt}}
\put(198,530){\makebox(0,0)[r]{0.8}}
\put(965.0,530.0){\rule[-0.500pt]{4.818pt}{1.000pt}}
\put(220.0,614.0){\rule[-0.500pt]{4.818pt}{1.000pt}}
\put(198,614){\makebox(0,0)[r]{0.9}}
\put(965.0,614.0){\rule[-0.500pt]{4.818pt}{1.000pt}}
\put(220.0,697.0){\rule[-0.500pt]{4.818pt}{1.000pt}}
\put(198,697){\makebox(0,0)[r]{1}}
\put(965.0,697.0){\rule[-0.500pt]{4.818pt}{1.000pt}}
\put(220.0,113.0){\rule[-0.500pt]{1.000pt}{4.818pt}}
\put(220,68){\makebox(0,0){0}}
\put(220.0,677.0){\rule[-0.500pt]{1.000pt}{4.818pt}}
\put(297.0,113.0){\rule[-0.500pt]{1.000pt}{4.818pt}}
\put(297,68){\makebox(0,0){0.1}}
\put(297.0,677.0){\rule[-0.500pt]{1.000pt}{4.818pt}}
\put(373.0,113.0){\rule[-0.500pt]{1.000pt}{4.818pt}}
\put(373,68){\makebox(0,0){0.2}}
\put(373.0,677.0){\rule[-0.500pt]{1.000pt}{4.818pt}}
\put(450.0,113.0){\rule[-0.500pt]{1.000pt}{4.818pt}}
\put(450,68){\makebox(0,0){0.3}}
\put(450.0,677.0){\rule[-0.500pt]{1.000pt}{4.818pt}}
\put(526.0,113.0){\rule[-0.500pt]{1.000pt}{4.818pt}}
\put(526,68){\makebox(0,0){0.4}}
\put(526.0,677.0){\rule[-0.500pt]{1.000pt}{4.818pt}}
\put(603.0,113.0){\rule[-0.500pt]{1.000pt}{4.818pt}}
\put(603,68){\makebox(0,0){0.5}}
\put(603.0,677.0){\rule[-0.500pt]{1.000pt}{4.818pt}}
\put(679.0,113.0){\rule[-0.500pt]{1.000pt}{4.818pt}}
\put(679,68){\makebox(0,0){0.6}}
\put(679.0,677.0){\rule[-0.500pt]{1.000pt}{4.818pt}}
\put(755.0,113.0){\rule[-0.500pt]{1.000pt}{4.818pt}}
\put(755,68){\makebox(0,0){0.7}}
\put(755.0,677.0){\rule[-0.500pt]{1.000pt}{4.818pt}}
\put(832.0,113.0){\rule[-0.500pt]{1.000pt}{4.818pt}}
\put(832,68){\makebox(0,0){0.8}}
\put(832.0,677.0){\rule[-0.500pt]{1.000pt}{4.818pt}}
\put(908.0,113.0){\rule[-0.500pt]{1.000pt}{4.818pt}}
\put(908,68){\makebox(0,0){0.9}}
\put(908.0,677.0){\rule[-0.500pt]{1.000pt}{4.818pt}}
\put(985.0,113.0){\rule[-0.500pt]{1.000pt}{4.818pt}}
\put(985,68){\makebox(0,0){1}}
\put(985.0,677.0){\rule[-0.500pt]{1.000pt}{4.818pt}}
\put(220.0,113.0){\rule[-0.500pt]{184.288pt}{1.000pt}}
\put(985.0,113.0){\rule[-0.500pt]{1.000pt}{140.686pt}}
\put(220.0,697.0){\rule[-0.500pt]{184.288pt}{1.000pt}}
\put(45,405){\makebox(0,0){{\cal F}}}
\put(602,23){\makebox(0,0){$E_0$}}
\put(220.0,113.0){\rule[-0.500pt]{1.000pt}{140.686pt}}
\put(855,632){\makebox(0,0)[r]{a}}
\put(899,632){\rule{1pt}{1pt}}
\put(984,697){\rule{1pt}{1pt}}
\put(773,640){\rule{1pt}{1pt}}
\put(691,594){\rule{1pt}{1pt}}
\put(631,554){\rule{1pt}{1pt}}
\put(584,521){\rule{1pt}{1pt}}
\put(545,491){\rule{1pt}{1pt}}
\put(512,465){\rule{1pt}{1pt}}
\put(483,442){\rule{1pt}{1pt}}
\put(457,421){\rule{1pt}{1pt}}
\put(435,403){\rule{1pt}{1pt}}
\put(415,386){\rule{1pt}{1pt}}
\put(397,370){\rule{1pt}{1pt}}
\put(381,356){\rule{1pt}{1pt}}
\put(366,342){\rule{1pt}{1pt}}
\put(352,330){\rule{1pt}{1pt}}
\put(340,319){\rule{1pt}{1pt}}
\put(329,308){\rule{1pt}{1pt}}
\put(319,298){\rule{1pt}{1pt}}
\put(310,289){\rule{1pt}{1pt}}
\put(302,280){\rule{1pt}{1pt}}
\put(294,271){\rule{1pt}{1pt}}
\put(287,263){\rule{1pt}{1pt}}
\put(280,256){\rule{1pt}{1pt}}
\put(274,249){\rule{1pt}{1pt}}
\put(269,242){\rule{1pt}{1pt}}
\put(264,235){\rule{1pt}{1pt}}
\put(259,229){\rule{1pt}{1pt}}
\put(255,223){\rule{1pt}{1pt}}
\put(251,218){\rule{1pt}{1pt}}
\put(248,212){\rule{1pt}{1pt}}
\put(244,207){\rule{1pt}{1pt}}
\put(241,202){\rule{1pt}{1pt}}
\put(239,197){\rule{1pt}{1pt}}
\put(236,192){\rule{1pt}{1pt}}
\put(234,188){\rule{1pt}{1pt}}
\put(232,184){\rule{1pt}{1pt}}
\put(230,179){\rule{1pt}{1pt}}
\put(229,175){\rule{1pt}{1pt}}
\put(227,172){\rule{1pt}{1pt}}
\put(226,168){\rule{1pt}{1pt}}
\put(225,164){\rule{1pt}{1pt}}
\put(224,161){\rule{1pt}{1pt}}
\put(223,157){\rule{1pt}{1pt}}
\put(222,154){\rule{1pt}{1pt}}
\put(222,151){\rule{1pt}{1pt}}
\put(221,148){\rule{1pt}{1pt}}
\put(221,144){\rule{1pt}{1pt}}
\put(220,141){\rule{1pt}{1pt}}
\put(220,139){\rule{1pt}{1pt}}
\put(220,136){\rule{1pt}{1pt}}
\put(855,587){\makebox(0,0)[r]{b}}
\put(877.0,587.0){\rule[-0.500pt]{15.899pt}{1.000pt}}
\put(984,697){\usebox{\plotpoint}}
\multiput(873.48,694.68)(-14.354,-0.481){8}{\rule{26.625pt}{0.116pt}}
\multiput(928.74,694.92)(-155.738,-8.000){2}{\rule{13.313pt}{1.000pt}}
\multiput(751.94,686.68)(-2.438,-0.494){26}{\rule{5.074pt}{0.119pt}}
\multiput(762.47,686.92)(-71.470,-17.000){2}{\rule{2.537pt}{1.000pt}}
\multiput(677.51,669.68)(-1.500,-0.495){32}{\rule{3.250pt}{0.119pt}}
\multiput(684.25,669.92)(-53.254,-20.000){2}{\rule{1.625pt}{1.000pt}}
\multiput(620.21,649.68)(-1.168,-0.495){32}{\rule{2.600pt}{0.119pt}}
\multiput(625.60,649.92)(-41.604,-20.000){2}{\rule{1.300pt}{1.000pt}}
\multiput(574.44,629.68)(-1.015,-0.495){30}{\rule{2.303pt}{0.119pt}}
\multiput(579.22,629.92)(-34.221,-19.000){2}{\rule{1.151pt}{1.000pt}}
\multiput(536.75,610.68)(-0.854,-0.495){30}{\rule{1.987pt}{0.119pt}}
\multiput(540.88,610.92)(-28.876,-19.000){2}{\rule{0.993pt}{1.000pt}}
\multiput(504.27,591.68)(-0.788,-0.495){28}{\rule{1.861pt}{0.119pt}}
\multiput(508.14,591.92)(-25.137,-18.000){2}{\rule{0.931pt}{1.000pt}}
\multiput(475.61,573.68)(-0.745,-0.494){26}{\rule{1.779pt}{0.119pt}}
\multiput(479.31,573.92)(-22.307,-17.000){2}{\rule{0.890pt}{1.000pt}}
\multiput(450.59,556.68)(-0.624,-0.494){26}{\rule{1.544pt}{0.119pt}}
\multiput(453.80,556.92)(-18.795,-17.000){2}{\rule{0.772pt}{1.000pt}}
\multiput(428.43,539.68)(-0.639,-0.493){22}{\rule{1.583pt}{0.119pt}}
\multiput(431.71,539.92)(-16.714,-15.000){2}{\rule{0.792pt}{1.000pt}}
\multiput(408.63,524.68)(-0.612,-0.492){20}{\rule{1.536pt}{0.119pt}}
\multiput(411.81,524.92)(-14.813,-14.000){2}{\rule{0.768pt}{1.000pt}}
\multiput(391.22,510.68)(-0.538,-0.492){20}{\rule{1.393pt}{0.119pt}}
\multiput(394.11,510.92)(-13.109,-14.000){2}{\rule{0.696pt}{1.000pt}}
\multiput(375.17,496.68)(-0.540,-0.492){18}{\rule{1.404pt}{0.118pt}}
\multiput(378.09,496.92)(-12.086,-13.000){2}{\rule{0.702pt}{1.000pt}}
\multiput(360.12,483.68)(-0.541,-0.491){16}{\rule{1.417pt}{0.118pt}}
\multiput(363.06,483.92)(-11.060,-12.000){2}{\rule{0.708pt}{1.000pt}}
\multiput(346.81,471.68)(-0.454,-0.491){16}{\rule{1.250pt}{0.118pt}}
\multiput(349.41,471.92)(-9.406,-12.000){2}{\rule{0.625pt}{1.000pt}}
\multiput(334.81,459.68)(-0.447,-0.489){14}{\rule{1.250pt}{0.118pt}}
\multiput(337.41,459.92)(-8.406,-11.000){2}{\rule{0.625pt}{1.000pt}}
\multiput(323.81,448.68)(-0.437,-0.487){12}{\rule{1.250pt}{0.117pt}}
\multiput(326.41,448.92)(-7.406,-10.000){2}{\rule{0.625pt}{1.000pt}}
\multiput(316.68,435.35)(-0.485,-0.483){10}{\rule{0.117pt}{1.361pt}}
\multiput(316.92,438.17)(-9.000,-7.175){2}{\rule{1.000pt}{0.681pt}}
\multiput(307.68,424.77)(-0.481,-0.539){8}{\rule{0.116pt}{1.500pt}}
\multiput(307.92,427.89)(-8.000,-6.887){2}{\rule{1.000pt}{0.750pt}}
\multiput(299.68,415.29)(-0.481,-0.470){8}{\rule{0.116pt}{1.375pt}}
\multiput(299.92,418.15)(-8.000,-6.146){2}{\rule{1.000pt}{0.688pt}}
\multiput(291.69,405.63)(-0.475,-0.525){6}{\rule{0.114pt}{1.536pt}}
\multiput(291.92,408.81)(-7.000,-5.813){2}{\rule{1.000pt}{0.768pt}}
\multiput(284.69,397.22)(-0.475,-0.444){6}{\rule{0.114pt}{1.393pt}}
\multiput(284.92,400.11)(-7.000,-5.109){2}{\rule{1.000pt}{0.696pt}}
\multiput(277.69,388.43)(-0.462,-0.476){4}{\rule{0.111pt}{1.583pt}}
\multiput(277.92,391.71)(-6.000,-4.714){2}{\rule{1.000pt}{0.792pt}}
\multiput(271.71,379.32)(-0.424,-0.320){2}{\rule{0.102pt}{1.850pt}}
\multiput(271.92,383.16)(-5.000,-4.160){2}{\rule{1.000pt}{0.925pt}}
\multiput(266.71,372.15)(-0.424,-0.151){2}{\rule{0.102pt}{1.650pt}}
\multiput(266.92,375.58)(-5.000,-3.575){2}{\rule{1.000pt}{0.825pt}}
\multiput(261.71,365.15)(-0.424,-0.151){2}{\rule{0.102pt}{1.650pt}}
\multiput(261.92,368.58)(-5.000,-3.575){2}{\rule{1.000pt}{0.825pt}}
\put(254.92,358){\rule{1.000pt}{1.686pt}}
\multiput(256.92,361.50)(-4.000,-3.500){2}{\rule{1.000pt}{0.843pt}}
\put(250.92,351){\rule{1.000pt}{1.686pt}}
\multiput(252.92,354.50)(-4.000,-3.500){2}{\rule{1.000pt}{0.843pt}}
\put(247.42,345){\rule{1.000pt}{1.445pt}}
\multiput(248.92,348.00)(-3.000,-3.000){2}{\rule{1.000pt}{0.723pt}}
\put(243.92,339){\rule{1.000pt}{1.445pt}}
\multiput(245.92,342.00)(-4.000,-3.000){2}{\rule{1.000pt}{0.723pt}}
\put(240.42,333){\rule{1.000pt}{1.445pt}}
\multiput(241.92,336.00)(-3.000,-3.000){2}{\rule{1.000pt}{0.723pt}}
\put(237.92,327){\rule{1.000pt}{1.445pt}}
\multiput(238.92,330.00)(-2.000,-3.000){2}{\rule{1.000pt}{0.723pt}}
\put(235.42,322){\rule{1.000pt}{1.204pt}}
\multiput(236.92,324.50)(-3.000,-2.500){2}{\rule{1.000pt}{0.602pt}}
\put(232.92,316){\rule{1.000pt}{1.445pt}}
\multiput(233.92,319.00)(-2.000,-3.000){2}{\rule{1.000pt}{0.723pt}}
\put(230.92,311){\rule{1.000pt}{1.204pt}}
\multiput(231.92,313.50)(-2.000,-2.500){2}{\rule{1.000pt}{0.602pt}}
\put(228.92,306){\rule{1.000pt}{1.204pt}}
\multiput(229.92,308.50)(-2.000,-2.500){2}{\rule{1.000pt}{0.602pt}}
\put(227.42,301){\rule{1.000pt}{1.204pt}}
\multiput(227.92,303.50)(-1.000,-2.500){2}{\rule{1.000pt}{0.602pt}}
\put(225.92,296){\rule{1.000pt}{1.204pt}}
\multiput(226.92,298.50)(-2.000,-2.500){2}{\rule{1.000pt}{0.602pt}}
\put(224.42,292){\rule{1.000pt}{0.964pt}}
\multiput(224.92,294.00)(-1.000,-2.000){2}{\rule{1.000pt}{0.482pt}}
\put(223.42,287){\rule{1.000pt}{1.204pt}}
\multiput(223.92,289.50)(-1.000,-2.500){2}{\rule{1.000pt}{0.602pt}}
\put(222.42,283){\rule{1.000pt}{0.964pt}}
\multiput(222.92,285.00)(-1.000,-2.000){2}{\rule{1.000pt}{0.482pt}}
\put(221.42,279){\rule{1.000pt}{0.964pt}}
\multiput(221.92,281.00)(-1.000,-2.000){2}{\rule{1.000pt}{0.482pt}}
\put(220.42,275){\rule{1.000pt}{0.964pt}}
\multiput(220.92,277.00)(-1.000,-2.000){2}{\rule{1.000pt}{0.482pt}}
\put(219.42,267){\rule{1.000pt}{0.964pt}}
\multiput(219.92,269.00)(-1.000,-2.000){2}{\rule{1.000pt}{0.482pt}}
\put(222.0,271.0){\usebox{\plotpoint}}
\put(218.42,259){\rule{1.000pt}{0.964pt}}
\multiput(218.92,261.00)(-1.000,-2.000){2}{\rule{1.000pt}{0.482pt}}
\put(221.0,263.0){\usebox{\plotpoint}}
\put(220.0,252.0){\rule[-0.500pt]{1.000pt}{1.686pt}}
\sbox{\plotpoint}{\rule[-0.175pt]{0.350pt}{0.350pt}}%
\put(855,542){\makebox(0,0)[r]{c}}
\put(877.0,542.0){\rule[-0.175pt]{15.899pt}{0.350pt}}
\put(984,697){\usebox{\plotpoint}}
\put(773,695.27){\rule{37.013pt}{0.350pt}}
\multiput(907.18,696.27)(-134.179,-2.000){2}{\rule{18.506pt}{0.350pt}}
\multiput(748.81,694.02)(-9.635,-0.507){7}{\rule{5.827pt}{0.122pt}}
\multiput(760.90,694.27)(-69.905,-5.000){2}{\rule{2.914pt}{0.350pt}}
\multiput(679.74,689.02)(-4.062,-0.504){13}{\rule{2.712pt}{0.121pt}}
\multiput(685.37,689.27)(-54.370,-8.000){2}{\rule{1.356pt}{0.350pt}}
\multiput(622.10,681.02)(-3.175,-0.504){13}{\rule{2.144pt}{0.121pt}}
\multiput(626.55,681.27)(-42.551,-8.000){2}{\rule{1.072pt}{0.350pt}}
\multiput(577.34,673.02)(-2.310,-0.503){15}{\rule{1.604pt}{0.121pt}}
\multiput(580.67,673.27)(-35.670,-9.000){2}{\rule{0.802pt}{0.350pt}}
\multiput(539.84,664.02)(-1.740,-0.503){17}{\rule{1.242pt}{0.121pt}}
\multiput(542.42,664.27)(-30.421,-10.000){2}{\rule{0.621pt}{0.350pt}}
\multiput(507.42,654.02)(-1.526,-0.503){17}{\rule{1.102pt}{0.121pt}}
\multiput(509.71,654.27)(-26.712,-10.000){2}{\rule{0.551pt}{0.350pt}}
\multiput(478.86,644.02)(-1.366,-0.503){17}{\rule{0.998pt}{0.121pt}}
\multiput(480.93,644.27)(-23.930,-10.000){2}{\rule{0.499pt}{0.350pt}}
\multiput(453.44,634.02)(-1.153,-0.503){17}{\rule{0.857pt}{0.121pt}}
\multiput(455.22,634.27)(-20.220,-10.000){2}{\rule{0.429pt}{0.350pt}}
\multiput(431.73,624.02)(-1.046,-0.503){17}{\rule{0.787pt}{0.121pt}}
\multiput(433.37,624.27)(-18.366,-10.000){2}{\rule{0.394pt}{0.350pt}}
\multiput(412.02,614.02)(-0.939,-0.503){17}{\rule{0.718pt}{0.121pt}}
\multiput(413.51,614.27)(-16.511,-10.000){2}{\rule{0.359pt}{0.350pt}}
\multiput(394.05,604.02)(-0.932,-0.503){15}{\rule{0.710pt}{0.121pt}}
\multiput(395.53,604.27)(-14.527,-9.000){2}{\rule{0.355pt}{0.350pt}}
\multiput(378.46,595.02)(-0.779,-0.503){17}{\rule{0.612pt}{0.121pt}}
\multiput(379.73,595.27)(-13.729,-10.000){2}{\rule{0.306pt}{0.350pt}}
\multiput(363.38,585.02)(-0.813,-0.503){15}{\rule{0.632pt}{0.121pt}}
\multiput(364.69,585.27)(-12.688,-9.000){2}{\rule{0.316pt}{0.350pt}}
\multiput(349.70,576.02)(-0.693,-0.503){15}{\rule{0.554pt}{0.121pt}}
\multiput(350.85,576.27)(-10.850,-9.000){2}{\rule{0.277pt}{0.350pt}}
\multiput(337.64,567.02)(-0.720,-0.504){13}{\rule{0.569pt}{0.121pt}}
\multiput(338.82,567.27)(-9.820,-8.000){2}{\rule{0.284pt}{0.350pt}}
\multiput(327.02,559.02)(-0.573,-0.503){15}{\rule{0.476pt}{0.121pt}}
\multiput(328.01,559.27)(-9.011,-9.000){2}{\rule{0.238pt}{0.350pt}}
\multiput(317.00,550.02)(-0.584,-0.504){13}{\rule{0.481pt}{0.121pt}}
\multiput(318.00,550.27)(-8.001,-8.000){2}{\rule{0.241pt}{0.350pt}}
\multiput(308.18,542.02)(-0.515,-0.504){13}{\rule{0.438pt}{0.121pt}}
\multiput(309.09,542.27)(-7.092,-8.000){2}{\rule{0.219pt}{0.350pt}}
\multiput(299.98,534.02)(-0.598,-0.504){11}{\rule{0.488pt}{0.121pt}}
\multiput(300.99,534.27)(-6.988,-7.000){2}{\rule{0.244pt}{0.350pt}}
\multiput(293.02,525.98)(-0.504,-0.598){11}{\rule{0.121pt}{0.488pt}}
\multiput(293.27,526.99)(-7.000,-6.988){2}{\rule{0.350pt}{0.244pt}}
\multiput(285.18,519.02)(-0.518,-0.504){11}{\rule{0.438pt}{0.121pt}}
\multiput(286.09,519.27)(-6.092,-7.000){2}{\rule{0.219pt}{0.350pt}}
\multiput(279.02,510.94)(-0.505,-0.618){9}{\rule{0.122pt}{0.496pt}}
\multiput(279.27,511.97)(-6.000,-5.971){2}{\rule{0.350pt}{0.248pt}}
\multiput(273.02,503.60)(-0.507,-0.767){7}{\rule{0.122pt}{0.577pt}}
\multiput(273.27,504.80)(-5.000,-5.801){2}{\rule{0.350pt}{0.289pt}}
\multiput(268.02,496.60)(-0.507,-0.767){7}{\rule{0.122pt}{0.577pt}}
\multiput(268.27,497.80)(-5.000,-5.801){2}{\rule{0.350pt}{0.289pt}}
\multiput(263.02,489.89)(-0.507,-0.649){7}{\rule{0.122pt}{0.507pt}}
\multiput(263.27,490.95)(-5.000,-4.947){2}{\rule{0.350pt}{0.254pt}}
\multiput(258.02,483.09)(-0.509,-1.024){5}{\rule{0.123pt}{0.700pt}}
\multiput(258.27,484.55)(-4.000,-5.547){2}{\rule{0.350pt}{0.350pt}}
\multiput(254.02,476.46)(-0.509,-0.864){5}{\rule{0.123pt}{0.613pt}}
\multiput(254.27,477.73)(-4.000,-4.729){2}{\rule{0.350pt}{0.306pt}}
\multiput(250.02,469.73)(-0.516,-1.366){3}{\rule{0.124pt}{0.787pt}}
\multiput(250.27,471.37)(-3.000,-4.366){2}{\rule{0.350pt}{0.394pt}}
\multiput(247.02,464.46)(-0.509,-0.864){5}{\rule{0.123pt}{0.613pt}}
\multiput(247.27,465.73)(-4.000,-4.729){2}{\rule{0.350pt}{0.306pt}}
\multiput(243.02,458.22)(-0.516,-1.108){3}{\rule{0.124pt}{0.671pt}}
\multiput(243.27,459.61)(-3.000,-3.608){2}{\rule{0.350pt}{0.335pt}}
\put(239.27,450){\rule{0.350pt}{1.137pt}}
\multiput(240.27,453.64)(-2.000,-3.639){2}{\rule{0.350pt}{0.569pt}}
\multiput(238.02,447.22)(-0.516,-1.108){3}{\rule{0.124pt}{0.671pt}}
\multiput(238.27,448.61)(-3.000,-3.608){2}{\rule{0.350pt}{0.335pt}}
\put(234.27,440){\rule{0.350pt}{0.962pt}}
\multiput(235.27,443.00)(-2.000,-3.002){2}{\rule{0.350pt}{0.481pt}}
\put(232.27,434){\rule{0.350pt}{1.137pt}}
\multiput(233.27,437.64)(-2.000,-3.639){2}{\rule{0.350pt}{0.569pt}}
\put(230.27,429){\rule{0.350pt}{0.962pt}}
\multiput(231.27,432.00)(-2.000,-3.002){2}{\rule{0.350pt}{0.481pt}}
\put(228.77,425){\rule{0.350pt}{0.964pt}}
\multiput(229.27,427.00)(-1.000,-2.000){2}{\rule{0.350pt}{0.482pt}}
\put(227.27,420){\rule{0.350pt}{0.962pt}}
\multiput(228.27,423.00)(-2.000,-3.002){2}{\rule{0.350pt}{0.481pt}}
\put(225.77,415){\rule{0.350pt}{1.204pt}}
\multiput(226.27,417.50)(-1.000,-2.500){2}{\rule{0.350pt}{0.602pt}}
\put(224.77,411){\rule{0.350pt}{0.964pt}}
\multiput(225.27,413.00)(-1.000,-2.000){2}{\rule{0.350pt}{0.482pt}}
\put(223.77,406){\rule{0.350pt}{1.204pt}}
\multiput(224.27,408.50)(-1.000,-2.500){2}{\rule{0.350pt}{0.602pt}}
\put(222.77,402){\rule{0.350pt}{0.964pt}}
\multiput(223.27,404.00)(-1.000,-2.000){2}{\rule{0.350pt}{0.482pt}}
\put(221.77,397){\rule{0.350pt}{1.204pt}}
\multiput(222.27,399.50)(-1.000,-2.500){2}{\rule{0.350pt}{0.602pt}}
\put(220.77,389){\rule{0.350pt}{0.964pt}}
\multiput(221.27,391.00)(-1.000,-2.000){2}{\rule{0.350pt}{0.482pt}}
\put(222.0,393.0){\rule[-0.175pt]{0.350pt}{0.964pt}}
\put(219.77,381){\rule{0.350pt}{0.964pt}}
\multiput(220.27,383.00)(-1.000,-2.000){2}{\rule{0.350pt}{0.482pt}}
\put(221.0,385.0){\rule[-0.175pt]{0.350pt}{0.964pt}}
\put(220.0,374.0){\rule[-0.175pt]{0.350pt}{1.686pt}}
\sbox{\plotpoint}{\rule[-0.300pt]{0.600pt}{0.600pt}}%
\put(855,497){\makebox(0,0)[r]{d}}
\put(877.0,497.0){\rule[-0.300pt]{15.899pt}{0.600pt}}
\put(984,697){\usebox{\plotpoint}}
\put(631,695.25){\rule{14.454pt}{0.600pt}}
\multiput(661.00,695.75)(-30.000,-1.000){2}{\rule{7.227pt}{0.600pt}}
\put(691.0,697.0){\rule[-0.300pt]{70.584pt}{0.600pt}}
\put(545,694.25){\rule{9.395pt}{0.600pt}}
\multiput(564.50,694.75)(-19.500,-1.000){2}{\rule{4.698pt}{0.600pt}}
\put(584.0,696.0){\rule[-0.300pt]{11.322pt}{0.600pt}}
\put(483,693.25){\rule{6.986pt}{0.600pt}}
\multiput(497.50,693.75)(-14.500,-1.000){2}{\rule{3.493pt}{0.600pt}}
\put(457,692.25){\rule{6.263pt}{0.600pt}}
\multiput(470.00,692.75)(-13.000,-1.000){2}{\rule{3.132pt}{0.600pt}}
\put(435,691.25){\rule{5.300pt}{0.600pt}}
\multiput(446.00,691.75)(-11.000,-1.000){2}{\rule{2.650pt}{0.600pt}}
\put(415,690.25){\rule{4.818pt}{0.600pt}}
\multiput(425.00,690.75)(-10.000,-1.000){2}{\rule{2.409pt}{0.600pt}}
\put(397,689.25){\rule{4.336pt}{0.600pt}}
\multiput(406.00,689.75)(-9.000,-1.000){2}{\rule{2.168pt}{0.600pt}}
\put(381,688.25){\rule{3.854pt}{0.600pt}}
\multiput(389.00,688.75)(-8.000,-1.000){2}{\rule{1.927pt}{0.600pt}}
\put(366,687.25){\rule{3.614pt}{0.600pt}}
\multiput(373.50,687.75)(-7.500,-1.000){2}{\rule{1.807pt}{0.600pt}}
\put(352,685.75){\rule{3.373pt}{0.600pt}}
\multiput(359.00,686.75)(-7.000,-2.000){2}{\rule{1.686pt}{0.600pt}}
\put(340,684.25){\rule{2.891pt}{0.600pt}}
\multiput(346.00,684.75)(-6.000,-1.000){2}{\rule{1.445pt}{0.600pt}}
\put(329,683.25){\rule{2.650pt}{0.600pt}}
\multiput(334.50,683.75)(-5.500,-1.000){2}{\rule{1.325pt}{0.600pt}}
\put(319,681.75){\rule{2.409pt}{0.600pt}}
\multiput(324.00,682.75)(-5.000,-2.000){2}{\rule{1.204pt}{0.600pt}}
\put(310,680.25){\rule{2.168pt}{0.600pt}}
\multiput(314.50,680.75)(-4.500,-1.000){2}{\rule{1.084pt}{0.600pt}}
\put(302,679.25){\rule{1.927pt}{0.600pt}}
\multiput(306.00,679.75)(-4.000,-1.000){2}{\rule{0.964pt}{0.600pt}}
\put(294,677.75){\rule{1.927pt}{0.600pt}}
\multiput(298.00,678.75)(-4.000,-2.000){2}{\rule{0.964pt}{0.600pt}}
\put(287,676.25){\rule{1.686pt}{0.600pt}}
\multiput(290.50,676.75)(-3.500,-1.000){2}{\rule{0.843pt}{0.600pt}}
\put(280,674.75){\rule{1.686pt}{0.600pt}}
\multiput(283.50,675.75)(-3.500,-2.000){2}{\rule{0.843pt}{0.600pt}}
\put(274,672.75){\rule{1.445pt}{0.600pt}}
\multiput(277.00,673.75)(-3.000,-2.000){2}{\rule{0.723pt}{0.600pt}}
\put(269,671.25){\rule{1.204pt}{0.600pt}}
\multiput(271.50,671.75)(-2.500,-1.000){2}{\rule{0.602pt}{0.600pt}}
\put(264,669.75){\rule{1.204pt}{0.600pt}}
\multiput(266.50,670.75)(-2.500,-2.000){2}{\rule{0.602pt}{0.600pt}}
\put(259,668.25){\rule{1.204pt}{0.600pt}}
\multiput(261.50,668.75)(-2.500,-1.000){2}{\rule{0.602pt}{0.600pt}}
\put(255,666.75){\rule{0.964pt}{0.600pt}}
\multiput(257.00,667.75)(-2.000,-2.000){2}{\rule{0.482pt}{0.600pt}}
\put(251,664.75){\rule{0.964pt}{0.600pt}}
\multiput(253.00,665.75)(-2.000,-2.000){2}{\rule{0.482pt}{0.600pt}}
\put(248,663.25){\rule{0.723pt}{0.600pt}}
\multiput(249.50,663.75)(-1.500,-1.000){2}{\rule{0.361pt}{0.600pt}}
\put(244,661.75){\rule{0.964pt}{0.600pt}}
\multiput(246.00,662.75)(-2.000,-2.000){2}{\rule{0.482pt}{0.600pt}}
\put(241,659.75){\rule{0.723pt}{0.600pt}}
\multiput(242.50,660.75)(-1.500,-2.000){2}{\rule{0.361pt}{0.600pt}}
\put(239,657.75){\rule{0.482pt}{0.600pt}}
\multiput(240.00,658.75)(-1.000,-2.000){2}{\rule{0.241pt}{0.600pt}}
\put(236,656.25){\rule{0.723pt}{0.600pt}}
\multiput(237.50,656.75)(-1.500,-1.000){2}{\rule{0.361pt}{0.600pt}}
\put(234,654.75){\rule{0.482pt}{0.600pt}}
\multiput(235.00,655.75)(-1.000,-2.000){2}{\rule{0.241pt}{0.600pt}}
\put(232,652.75){\rule{0.482pt}{0.600pt}}
\multiput(233.00,653.75)(-1.000,-2.000){2}{\rule{0.241pt}{0.600pt}}
\put(230,650.75){\rule{0.482pt}{0.600pt}}
\multiput(231.00,651.75)(-1.000,-2.000){2}{\rule{0.241pt}{0.600pt}}
\put(229,649.25){\rule{0.241pt}{0.600pt}}
\multiput(229.50,649.75)(-0.500,-1.000){2}{\rule{0.120pt}{0.600pt}}
\put(227,647.75){\rule{0.482pt}{0.600pt}}
\multiput(228.00,648.75)(-1.000,-2.000){2}{\rule{0.241pt}{0.600pt}}
\put(225.25,646){\rule{0.600pt}{0.482pt}}
\multiput(225.75,647.00)(-1.000,-1.000){2}{\rule{0.600pt}{0.241pt}}
\put(224.25,644){\rule{0.600pt}{0.482pt}}
\multiput(224.75,645.00)(-1.000,-1.000){2}{\rule{0.600pt}{0.241pt}}
\put(224,642.25){\rule{0.241pt}{0.600pt}}
\multiput(224.50,642.75)(-0.500,-1.000){2}{\rule{0.120pt}{0.600pt}}
\put(222.25,641){\rule{0.600pt}{0.482pt}}
\multiput(222.75,642.00)(-1.000,-1.000){2}{\rule{0.600pt}{0.241pt}}
\put(221.25,639){\rule{0.600pt}{0.482pt}}
\multiput(221.75,640.00)(-1.000,-1.000){2}{\rule{0.600pt}{0.241pt}}
\put(512.0,695.0){\rule[-0.300pt]{7.950pt}{0.600pt}}
\put(220.25,635){\rule{0.600pt}{0.482pt}}
\multiput(220.75,636.00)(-1.000,-1.000){2}{\rule{0.600pt}{0.241pt}}
\put(222.0,637.0){\usebox{\plotpoint}}
\put(219.25,632){\rule{0.600pt}{0.482pt}}
\multiput(219.75,633.00)(-1.000,-1.000){2}{\rule{0.600pt}{0.241pt}}
\put(221.0,634.0){\usebox{\plotpoint}}
\put(220.0,628.0){\rule[-0.300pt]{0.600pt}{0.964pt}}
\end{picture}

%% file: non.tex
\setlength{\unitlength}{0.240900pt}
\ifx\plotpoint\undefined\newsavebox{\plotpoint}\fi
\sbox{\plotpoint}{\rule[-0.500pt]{1.000pt}{1.000pt}}%
\begin{picture}(1049,720)(0,0)
\font\gnuplot=cmr10 at 10pt
\gnuplot
\sbox{\plotpoint}{\rule[-0.500pt]{1.000pt}{1.000pt}}%
\put(220.0,113.0){\rule[-0.500pt]{184.288pt}{1.000pt}}
\put(220.0,113.0){\rule[-0.500pt]{1.000pt}{140.686pt}}
\put(220.0,113.0){\rule[-0.500pt]{4.818pt}{1.000pt}}
\put(198,113){\makebox(0,0)[r]{0}}
\put(965.0,113.0){\rule[-0.500pt]{4.818pt}{1.000pt}}
\put(220.0,171.0){\rule[-0.500pt]{4.818pt}{1.000pt}}
\put(198,171){\makebox(0,0)[r]{0.1}}
\put(965.0,171.0){\rule[-0.500pt]{4.818pt}{1.000pt}}
\put(220.0,230.0){\rule[-0.500pt]{4.818pt}{1.000pt}}
\put(198,230){\makebox(0,0)[r]{0.2}}
\put(965.0,230.0){\rule[-0.500pt]{4.818pt}{1.000pt}}
\put(220.0,288.0){\rule[-0.500pt]{4.818pt}{1.000pt}}
\put(198,288){\makebox(0,0)[r]{0.3}}
\put(965.0,288.0){\rule[-0.500pt]{4.818pt}{1.000pt}}
\put(220.0,347.0){\rule[-0.500pt]{4.818pt}{1.000pt}}
\put(198,347){\makebox(0,0)[r]{0.4}}
\put(965.0,347.0){\rule[-0.500pt]{4.818pt}{1.000pt}}
\put(220.0,405.0){\rule[-0.500pt]{4.818pt}{1.000pt}}
\put(198,405){\makebox(0,0)[r]{0.5}}
\put(965.0,405.0){\rule[-0.500pt]{4.818pt}{1.000pt}}
\put(220.0,463.0){\rule[-0.500pt]{4.818pt}{1.000pt}}
\put(198,463){\makebox(0,0)[r]{0.6}}
\put(965.0,463.0){\rule[-0.500pt]{4.818pt}{1.000pt}}
\put(220.0,522.0){\rule[-0.500pt]{4.818pt}{1.000pt}}
\put(198,522){\makebox(0,0)[r]{0.7}}
\put(965.0,522.0){\rule[-0.500pt]{4.818pt}{1.000pt}}
\put(220.0,580.0){\rule[-0.500pt]{4.818pt}{1.000pt}}
\put(198,580){\makebox(0,0)[r]{0.8}}
\put(965.0,580.0){\rule[-0.500pt]{4.818pt}{1.000pt}}
\put(220.0,639.0){\rule[-0.500pt]{4.818pt}{1.000pt}}
\put(198,639){\makebox(0,0)[r]{0.9}}
\put(965.0,639.0){\rule[-0.500pt]{4.818pt}{1.000pt}}
\put(220.0,697.0){\rule[-0.500pt]{4.818pt}{1.000pt}}
\put(198,697){\makebox(0,0)[r]{1}}
\put(965.0,697.0){\rule[-0.500pt]{4.818pt}{1.000pt}}
\put(220.0,113.0){\rule[-0.500pt]{1.000pt}{4.818pt}}
\put(220,68){\makebox(0,0){0}}
\put(220.0,677.0){\rule[-0.500pt]{1.000pt}{4.818pt}}
\put(373.0,113.0){\rule[-0.500pt]{1.000pt}{4.818pt}}
\put(373,68){\makebox(0,0){0.2}}
\put(373.0,677.0){\rule[-0.500pt]{1.000pt}{4.818pt}}
\put(526.0,113.0){\rule[-0.500pt]{1.000pt}{4.818pt}}
\put(526,68){\makebox(0,0){0.4}}
\put(526.0,677.0){\rule[-0.500pt]{1.000pt}{4.818pt}}
\put(679.0,113.0){\rule[-0.500pt]{1.000pt}{4.818pt}}
\put(679,68){\makebox(0,0){0.6}}
\put(679.0,677.0){\rule[-0.500pt]{1.000pt}{4.818pt}}
\put(832.0,113.0){\rule[-0.500pt]{1.000pt}{4.818pt}}
\put(832,68){\makebox(0,0){0.8}}
\put(832.0,677.0){\rule[-0.500pt]{1.000pt}{4.818pt}}
\put(985.0,113.0){\rule[-0.500pt]{1.000pt}{4.818pt}}
\put(985,68){\makebox(0,0){1}}
\put(985.0,677.0){\rule[-0.500pt]{1.000pt}{4.818pt}}
\put(220.0,113.0){\rule[-0.500pt]{184.288pt}{1.000pt}}
\put(985.0,113.0){\rule[-0.500pt]{1.000pt}{140.686pt}}
\put(220.0,697.0){\rule[-0.500pt]{184.288pt}{1.000pt}}
\put(45,405){\makebox(0,0){$Q_{out}$}}
\put(602,23){\makebox(0,0){$Q_{in}$}}
\put(220.0,113.0){\rule[-0.500pt]{1.000pt}{140.686pt}}
\put(855,632){\makebox(0,0)[r]{a}}
\put(877.0,632.0){\rule[-0.500pt]{15.899pt}{1.000pt}}
\put(220,113){\usebox{\plotpoint}}
\put(220,113){\usebox{\plotpoint}}
\put(220,113){\usebox{\plotpoint}}
\put(220,113){\usebox{\plotpoint}}
\put(220,113){\usebox{\plotpoint}}
\put(220.0,113.0){\usebox{\plotpoint}}
\put(221.0,113.0){\usebox{\plotpoint}}
\put(221.0,114.0){\usebox{\plotpoint}}
\put(222.0,114.0){\usebox{\plotpoint}}
\put(223,113.42){\rule{0.241pt}{1.000pt}}
\multiput(223.00,112.92)(0.500,1.000){2}{\rule{0.120pt}{1.000pt}}
\put(222.0,115.0){\usebox{\plotpoint}}
\put(224,116){\usebox{\plotpoint}}
\put(224,114.42){\rule{0.241pt}{1.000pt}}
\multiput(224.00,113.92)(0.500,1.000){2}{\rule{0.120pt}{1.000pt}}
\put(225,115.42){\rule{0.241pt}{1.000pt}}
\multiput(225.00,114.92)(0.500,1.000){2}{\rule{0.120pt}{1.000pt}}
\put(227,116.42){\rule{0.241pt}{1.000pt}}
\multiput(227.00,115.92)(0.500,1.000){2}{\rule{0.120pt}{1.000pt}}
\put(228,117.42){\rule{0.241pt}{1.000pt}}
\multiput(228.00,116.92)(0.500,1.000){2}{\rule{0.120pt}{1.000pt}}
\put(229,118.42){\rule{0.482pt}{1.000pt}}
\multiput(229.00,117.92)(1.000,1.000){2}{\rule{0.241pt}{1.000pt}}
\put(231,119.42){\rule{0.241pt}{1.000pt}}
\multiput(231.00,118.92)(0.500,1.000){2}{\rule{0.120pt}{1.000pt}}
\put(232,120.92){\rule{0.482pt}{1.000pt}}
\multiput(232.00,119.92)(1.000,2.000){2}{\rule{0.241pt}{1.000pt}}
\put(234,122.42){\rule{0.482pt}{1.000pt}}
\multiput(234.00,121.92)(1.000,1.000){2}{\rule{0.241pt}{1.000pt}}
\put(236,123.42){\rule{0.482pt}{1.000pt}}
\multiput(236.00,122.92)(1.000,1.000){2}{\rule{0.241pt}{1.000pt}}
\put(238,124.92){\rule{0.482pt}{1.000pt}}
\multiput(238.00,123.92)(1.000,2.000){2}{\rule{0.241pt}{1.000pt}}
\put(240,126.92){\rule{0.482pt}{1.000pt}}
\multiput(240.00,125.92)(1.000,2.000){2}{\rule{0.241pt}{1.000pt}}
\put(242,128.92){\rule{0.723pt}{1.000pt}}
\multiput(242.00,127.92)(1.500,2.000){2}{\rule{0.361pt}{1.000pt}}
\put(245,130.92){\rule{0.723pt}{1.000pt}}
\multiput(245.00,129.92)(1.500,2.000){2}{\rule{0.361pt}{1.000pt}}
\put(248,132.92){\rule{0.723pt}{1.000pt}}
\multiput(248.00,131.92)(1.500,2.000){2}{\rule{0.361pt}{1.000pt}}
\put(251,135.42){\rule{0.723pt}{1.000pt}}
\multiput(251.00,133.92)(1.500,3.000){2}{\rule{0.361pt}{1.000pt}}
\put(254,138.42){\rule{0.964pt}{1.000pt}}
\multiput(254.00,136.92)(2.000,3.000){2}{\rule{0.482pt}{1.000pt}}
\put(258,141.42){\rule{0.964pt}{1.000pt}}
\multiput(258.00,139.92)(2.000,3.000){2}{\rule{0.482pt}{1.000pt}}
\put(262,144.42){\rule{0.964pt}{1.000pt}}
\multiput(262.00,142.92)(2.000,3.000){2}{\rule{0.482pt}{1.000pt}}
\put(266,147.92){\rule{1.204pt}{1.000pt}}
\multiput(266.00,145.92)(2.500,4.000){2}{\rule{0.602pt}{1.000pt}}
\put(271,151.92){\rule{1.445pt}{1.000pt}}
\multiput(271.00,149.92)(3.000,4.000){2}{\rule{0.723pt}{1.000pt}}
\put(277,156.42){\rule{1.445pt}{1.000pt}}
\multiput(277.00,153.92)(3.000,5.000){2}{\rule{0.723pt}{1.000pt}}
\put(283,161.42){\rule{1.445pt}{1.000pt}}
\multiput(283.00,158.92)(3.000,5.000){2}{\rule{0.723pt}{1.000pt}}
\multiput(289.00,167.84)(0.476,0.462){4}{\rule{1.583pt}{0.111pt}}
\multiput(289.00,163.92)(4.714,6.000){2}{\rule{0.792pt}{1.000pt}}
\multiput(297.00,173.84)(0.476,0.462){4}{\rule{1.583pt}{0.111pt}}
\multiput(297.00,169.92)(4.714,6.000){2}{\rule{0.792pt}{1.000pt}}
\multiput(305.00,179.84)(0.525,0.475){6}{\rule{1.536pt}{0.114pt}}
\multiput(305.00,175.92)(5.813,7.000){2}{\rule{0.768pt}{1.000pt}}
\multiput(314.00,186.83)(0.608,0.481){8}{\rule{1.625pt}{0.116pt}}
\multiput(314.00,182.92)(7.627,8.000){2}{\rule{0.813pt}{1.000pt}}
\multiput(325.00,194.83)(0.543,0.485){10}{\rule{1.472pt}{0.117pt}}
\multiput(325.00,190.92)(7.944,9.000){2}{\rule{0.736pt}{1.000pt}}
\multiput(336.00,203.83)(0.650,0.487){12}{\rule{1.650pt}{0.117pt}}
\multiput(336.00,199.92)(10.575,10.000){2}{\rule{0.825pt}{1.000pt}}
\multiput(350.00,213.83)(0.585,0.491){16}{\rule{1.500pt}{0.118pt}}
\multiput(350.00,209.92)(11.887,12.000){2}{\rule{0.750pt}{1.000pt}}
\multiput(365.00,225.83)(0.612,0.492){20}{\rule{1.536pt}{0.119pt}}
\multiput(365.00,221.92)(14.813,14.000){2}{\rule{0.768pt}{1.000pt}}
\multiput(383.00,239.83)(0.663,0.494){24}{\rule{1.625pt}{0.119pt}}
\multiput(383.00,235.92)(18.627,16.000){2}{\rule{0.813pt}{1.000pt}}
\multiput(405.00,255.83)(0.631,0.495){32}{\rule{1.550pt}{0.119pt}}
\multiput(405.00,251.92)(22.783,20.000){2}{\rule{0.775pt}{1.000pt}}
\multiput(431.00,275.83)(0.640,0.497){44}{\rule{1.558pt}{0.120pt}}
\multiput(431.00,271.92)(30.767,26.000){2}{\rule{0.779pt}{1.000pt}}
\multiput(465.00,301.83)(0.648,0.498){62}{\rule{1.564pt}{0.120pt}}
\multiput(465.00,297.92)(42.753,35.000){2}{\rule{0.782pt}{1.000pt}}
\multiput(511.00,336.83)(0.652,0.498){100}{\rule{1.565pt}{0.120pt}}
\multiput(511.00,332.92)(67.752,54.000){2}{\rule{0.782pt}{1.000pt}}
\multiput(582.00,390.83)(0.654,0.500){566}{\rule{1.560pt}{0.120pt}}
\multiput(582.00,386.92)(372.762,287.000){2}{\rule{0.780pt}{1.000pt}}
\put(226.0,118.0){\usebox{\plotpoint}}
\sbox{\plotpoint}{\rule[-0.175pt]{0.350pt}{0.350pt}}%
\put(855,587){\makebox(0,0)[r]{b}}
\put(877.0,587.0){\rule[-0.175pt]{15.899pt}{0.350pt}}
\put(220,113){\usebox{\plotpoint}}
\put(220.0,113.0){\rule[-0.175pt]{0.482pt}{0.350pt}}
\put(222.0,113.0){\usebox{\plotpoint}}
\put(224,113.77){\rule{0.241pt}{0.350pt}}
\multiput(224.00,113.27)(0.500,1.000){2}{\rule{0.120pt}{0.350pt}}
\put(222.0,114.0){\rule[-0.175pt]{0.482pt}{0.350pt}}
\put(226,114.77){\rule{0.241pt}{0.350pt}}
\multiput(226.00,114.27)(0.500,1.000){2}{\rule{0.120pt}{0.350pt}}
\put(225.0,115.0){\usebox{\plotpoint}}
\put(228,115.77){\rule{0.241pt}{0.350pt}}
\multiput(228.00,115.27)(0.500,1.000){2}{\rule{0.120pt}{0.350pt}}
\put(229,116.77){\rule{0.482pt}{0.350pt}}
\multiput(229.00,116.27)(1.000,1.000){2}{\rule{0.241pt}{0.350pt}}
\put(227.0,116.0){\usebox{\plotpoint}}
\put(232,117.77){\rule{0.482pt}{0.350pt}}
\multiput(232.00,117.27)(1.000,1.000){2}{\rule{0.241pt}{0.350pt}}
\put(234,118.77){\rule{0.482pt}{0.350pt}}
\multiput(234.00,118.27)(1.000,1.000){2}{\rule{0.241pt}{0.350pt}}
\put(236,119.77){\rule{0.482pt}{0.350pt}}
\multiput(236.00,119.27)(1.000,1.000){2}{\rule{0.241pt}{0.350pt}}
\put(238,120.77){\rule{0.482pt}{0.350pt}}
\multiput(238.00,120.27)(1.000,1.000){2}{\rule{0.241pt}{0.350pt}}
\put(240,122.27){\rule{0.438pt}{0.350pt}}
\multiput(240.00,121.27)(1.092,2.000){2}{\rule{0.219pt}{0.350pt}}
\put(242,123.77){\rule{0.723pt}{0.350pt}}
\multiput(242.00,123.27)(1.500,1.000){2}{\rule{0.361pt}{0.350pt}}
\put(245,124.77){\rule{0.723pt}{0.350pt}}
\multiput(245.00,124.27)(1.500,1.000){2}{\rule{0.361pt}{0.350pt}}
\put(248,126.27){\rule{0.612pt}{0.350pt}}
\multiput(248.00,125.27)(1.729,2.000){2}{\rule{0.306pt}{0.350pt}}
\put(251,128.27){\rule{0.612pt}{0.350pt}}
\multiput(251.00,127.27)(1.729,2.000){2}{\rule{0.306pt}{0.350pt}}
\put(254,130.27){\rule{0.787pt}{0.350pt}}
\multiput(254.00,129.27)(2.366,2.000){2}{\rule{0.394pt}{0.350pt}}
\put(258,132.27){\rule{0.787pt}{0.350pt}}
\multiput(258.00,131.27)(2.366,2.000){2}{\rule{0.394pt}{0.350pt}}
\put(262,134.27){\rule{0.787pt}{0.350pt}}
\multiput(262.00,133.27)(2.366,2.000){2}{\rule{0.394pt}{0.350pt}}
\multiput(266.00,136.47)(1.108,0.516){3}{\rule{0.671pt}{0.124pt}}
\multiput(266.00,135.27)(3.608,3.000){2}{\rule{0.335pt}{0.350pt}}
\multiput(271.00,139.47)(1.366,0.516){3}{\rule{0.787pt}{0.124pt}}
\multiput(271.00,138.27)(4.366,3.000){2}{\rule{0.394pt}{0.350pt}}
\multiput(277.00,142.47)(1.366,0.516){3}{\rule{0.787pt}{0.124pt}}
\multiput(277.00,141.27)(4.366,3.000){2}{\rule{0.394pt}{0.350pt}}
\multiput(283.00,145.47)(1.366,0.516){3}{\rule{0.787pt}{0.124pt}}
\multiput(283.00,144.27)(4.366,3.000){2}{\rule{0.394pt}{0.350pt}}
\multiput(289.00,148.47)(1.183,0.509){5}{\rule{0.787pt}{0.123pt}}
\multiput(289.00,147.27)(6.366,4.000){2}{\rule{0.394pt}{0.350pt}}
\multiput(297.00,152.47)(1.183,0.509){5}{\rule{0.787pt}{0.123pt}}
\multiput(297.00,151.27)(6.366,4.000){2}{\rule{0.394pt}{0.350pt}}
\multiput(305.00,156.47)(1.004,0.507){7}{\rule{0.718pt}{0.122pt}}
\multiput(305.00,155.27)(7.511,5.000){2}{\rule{0.359pt}{0.350pt}}
\multiput(314.00,161.47)(1.240,0.507){7}{\rule{0.857pt}{0.122pt}}
\multiput(314.00,160.27)(9.220,5.000){2}{\rule{0.429pt}{0.350pt}}
\multiput(325.00,166.47)(0.997,0.505){9}{\rule{0.729pt}{0.122pt}}
\multiput(325.00,165.27)(9.487,6.000){2}{\rule{0.365pt}{0.350pt}}
\multiput(336.00,172.47)(1.281,0.505){9}{\rule{0.904pt}{0.122pt}}
\multiput(336.00,171.27)(12.123,6.000){2}{\rule{0.452pt}{0.350pt}}
\multiput(350.00,178.47)(1.152,0.504){11}{\rule{0.837pt}{0.121pt}}
\multiput(350.00,177.27)(13.262,7.000){2}{\rule{0.419pt}{0.350pt}}
\multiput(365.00,185.47)(1.197,0.504){13}{\rule{0.875pt}{0.121pt}}
\multiput(365.00,184.27)(16.184,8.000){2}{\rule{0.438pt}{0.350pt}}
\multiput(383.00,193.47)(1.292,0.503){15}{\rule{0.943pt}{0.121pt}}
\multiput(383.00,192.27)(20.043,9.000){2}{\rule{0.472pt}{0.350pt}}
\multiput(405.00,202.48)(1.366,0.503){17}{\rule{0.998pt}{0.121pt}}
\multiput(405.00,201.27)(23.930,10.000){2}{\rule{0.499pt}{0.350pt}}
\multiput(431.00,212.48)(1.476,0.502){21}{\rule{1.079pt}{0.121pt}}
\multiput(431.00,211.27)(31.760,12.000){2}{\rule{0.540pt}{0.350pt}}
\multiput(465.00,224.48)(1.703,0.502){25}{\rule{1.238pt}{0.121pt}}
\multiput(465.00,223.27)(43.432,14.000){2}{\rule{0.619pt}{0.350pt}}
\multiput(511.00,238.48)(2.292,0.502){29}{\rule{1.641pt}{0.121pt}}
\multiput(511.00,237.27)(67.595,16.000){2}{\rule{0.820pt}{0.350pt}}
\multiput(582.00,254.48)(9.676,0.501){37}{\rule{6.668pt}{0.121pt}}
\multiput(582.00,253.27)(362.161,20.000){2}{\rule{3.334pt}{0.350pt}}
\put(231.0,118.0){\usebox{\plotpoint}}
\sbox{\plotpoint}{\rule[-0.300pt]{0.600pt}{0.600pt}}%
\put(855,542){\makebox(0,0)[r]{c}}
\put(877.0,542.0){\rule[-0.300pt]{15.899pt}{0.600pt}}
\put(222,113){\usebox{\plotpoint}}
\put(226,112.25){\rule{0.241pt}{0.600pt}}
\multiput(226.00,111.75)(0.500,1.000){2}{\rule{0.120pt}{0.600pt}}
\put(222.0,113.0){\rule[-0.300pt]{0.964pt}{0.600pt}}
\put(231,113.25){\rule{0.241pt}{0.600pt}}
\multiput(231.00,112.75)(0.500,1.000){2}{\rule{0.120pt}{0.600pt}}
\put(227.0,114.0){\rule[-0.300pt]{0.964pt}{0.600pt}}
\put(234,114.25){\rule{0.482pt}{0.600pt}}
\multiput(234.00,113.75)(1.000,1.000){2}{\rule{0.241pt}{0.600pt}}
\put(232.0,115.0){\usebox{\plotpoint}}
\put(238,115.25){\rule{0.482pt}{0.600pt}}
\multiput(238.00,114.75)(1.000,1.000){2}{\rule{0.241pt}{0.600pt}}
\put(240,116.25){\rule{0.482pt}{0.600pt}}
\multiput(240.00,115.75)(1.000,1.000){2}{\rule{0.241pt}{0.600pt}}
\put(236.0,116.0){\usebox{\plotpoint}}
\put(245,117.25){\rule{0.723pt}{0.600pt}}
\multiput(245.00,116.75)(1.500,1.000){2}{\rule{0.361pt}{0.600pt}}
\put(248,118.25){\rule{0.723pt}{0.600pt}}
\multiput(248.00,117.75)(1.500,1.000){2}{\rule{0.361pt}{0.600pt}}
\put(251,119.25){\rule{0.723pt}{0.600pt}}
\multiput(251.00,118.75)(1.500,1.000){2}{\rule{0.361pt}{0.600pt}}
\put(254,120.25){\rule{0.964pt}{0.600pt}}
\multiput(254.00,119.75)(2.000,1.000){2}{\rule{0.482pt}{0.600pt}}
\put(258,121.75){\rule{0.964pt}{0.600pt}}
\multiput(258.00,120.75)(2.000,2.000){2}{\rule{0.482pt}{0.600pt}}
\put(262,123.25){\rule{0.964pt}{0.600pt}}
\multiput(262.00,122.75)(2.000,1.000){2}{\rule{0.482pt}{0.600pt}}
\put(266,124.25){\rule{1.204pt}{0.600pt}}
\multiput(266.00,123.75)(2.500,1.000){2}{\rule{0.602pt}{0.600pt}}
\put(271,125.75){\rule{1.445pt}{0.600pt}}
\multiput(271.00,124.75)(3.000,2.000){2}{\rule{0.723pt}{0.600pt}}
\put(277,127.75){\rule{1.445pt}{0.600pt}}
\multiput(277.00,126.75)(3.000,2.000){2}{\rule{0.723pt}{0.600pt}}
\put(283,129.75){\rule{1.445pt}{0.600pt}}
\multiput(283.00,128.75)(3.000,2.000){2}{\rule{0.723pt}{0.600pt}}
\put(289,131.75){\rule{1.927pt}{0.600pt}}
\multiput(289.00,130.75)(4.000,2.000){2}{\rule{0.964pt}{0.600pt}}
\put(297,133.75){\rule{1.927pt}{0.600pt}}
\multiput(297.00,132.75)(4.000,2.000){2}{\rule{0.964pt}{0.600pt}}
\put(305,136.25){\rule{1.950pt}{0.600pt}}
\multiput(305.00,134.75)(4.953,3.000){2}{\rule{0.975pt}{0.600pt}}
\put(314,139.25){\rule{2.350pt}{0.600pt}}
\multiput(314.00,137.75)(6.122,3.000){2}{\rule{1.175pt}{0.600pt}}
\put(325,142.25){\rule{2.350pt}{0.600pt}}
\multiput(325.00,140.75)(6.122,3.000){2}{\rule{1.175pt}{0.600pt}}
\put(336,145.25){\rule{2.950pt}{0.600pt}}
\multiput(336.00,143.75)(7.877,3.000){2}{\rule{1.475pt}{0.600pt}}
\multiput(350.00,148.99)(2.519,0.503){3}{\rule{2.400pt}{0.121pt}}
\multiput(350.00,146.75)(10.019,4.000){2}{\rule{1.200pt}{0.600pt}}
\multiput(365.00,152.99)(3.085,0.503){3}{\rule{2.850pt}{0.121pt}}
\multiput(365.00,150.75)(12.085,4.000){2}{\rule{1.425pt}{0.600pt}}
\multiput(383.00,156.99)(2.605,0.502){5}{\rule{2.790pt}{0.121pt}}
\multiput(383.00,154.75)(16.209,5.000){2}{\rule{1.395pt}{0.600pt}}
\multiput(405.00,161.99)(3.106,0.502){5}{\rule{3.270pt}{0.121pt}}
\multiput(405.00,159.75)(19.213,5.000){2}{\rule{1.635pt}{0.600pt}}
\multiput(431.00,166.99)(3.211,0.501){7}{\rule{3.550pt}{0.121pt}}
\multiput(431.00,164.75)(26.632,6.000){2}{\rule{1.775pt}{0.600pt}}
\multiput(465.00,172.99)(4.380,0.501){7}{\rule{4.750pt}{0.121pt}}
\multiput(465.00,170.75)(36.141,6.000){2}{\rule{2.375pt}{0.600pt}}
\multiput(511.00,178.99)(5.639,0.501){9}{\rule{6.236pt}{0.121pt}}
\multiput(511.00,176.75)(58.057,7.000){2}{\rule{3.118pt}{0.600pt}}
\put(582,184.25){\rule{97.083pt}{0.600pt}}
\multiput(582.00,183.75)(201.500,1.000){2}{\rule{48.541pt}{0.600pt}}
\put(242.0,118.0){\rule[-0.300pt]{0.723pt}{0.600pt}}
\end{picture}